\PassOptionsToPackage{unicode}{hyperref}
\PassOptionsToPackage{hyphens}{url}
\documentclass[
]{article}
\usepackage{amsmath,amssymb}
\usepackage{iftex}
\ifPDFTeX
  \usepackage[T1]{fontenc}
  \usepackage[utf8]{inputenc}
  \usepackage{textcomp} 
\else 
  \usepackage{unicode-math} 
  \defaultfontfeatures{Scale=MatchLowercase}
  \defaultfontfeatures[\rmfamily]{Ligatures=TeX,Scale=1}
\fi
\usepackage{lmodern}
\ifPDFTeX\else
\fi
\IfFileExists{upquote.sty}{\usepackage{upquote}}{}
\IfFileExists{microtype.sty}{
  \usepackage[]{microtype}
  \UseMicrotypeSet[protrusion]{basicmath} 
}{}
\makeatletter
\@ifundefined{KOMAClassName}{
  \IfFileExists{parskip.sty}{%
    \usepackage{parskip}
  }{
    \setlength{\parindent}{0pt}
    \setlength{\parskip}{6pt plus 2pt minus 1pt}}
}{
  \KOMAoptions{parskip=half}}
\makeatother
\usepackage{xcolor}
\usepackage{longtable,booktabs,array}
\usepackage{calc} 
\usepackage{etoolbox}
\makeatletter
\patchcmd\longtable{\par}{\if@noskipsec\mbox{}\fi\par}{}{}
\makeatother
\IfFileExists{footnotehyper.sty}{\usepackage{footnotehyper}}{\usepackage{footnote}}
\makesavenoteenv{longtable}
\usepackage{graphicx}
\makeatletter
\def\maxwidth{\ifdim\Gin@nat@width>\linewidth\linewidth\else\Gin@nat@width\fi}
\def\maxheight{\ifdim\Gin@nat@height>\textheight\textheight\else\Gin@nat@height\fi}
\makeatother
\setkeys{Gin}{width=\maxwidth,height=\maxheight,keepaspectratio}
\makeatletter
\def\fps@figure{htbp}
\makeatother
\ifLuaTeX
  \usepackage{selnolig}  
\fi
\usepackage{bookmark}
\IfFileExists{xurl.sty}{\usepackage{xurl}}{} 
\hypersetup{
  hidelinks,
  pdfcreator={LaTeX via pandoc}}

\author{}
\date{}

\begin{document}

\textbf{Opening Rankings from Engine Self-Play Are Less Certain Than
They Look}

\emph{Sampling Error, Draws, and Strength Dependence in Strength-Limited
Stockfish Games}

Conrad Flick\textsuperscript{*}, Jaysen Jensen, Jingyu Huang, and Mario
Camara Paslar

University of Chicago, Chicago, Illinois, USA

\emph{Corresponding author: flickc@uchicago.edu}

Emails: flickc@uchicago.edu (C. Flick), jaysen@uchicago.edu (J. Jensen),
jyhuang28@uchicago.edu (J. Huang), MarioCamaraPaslar@gmail.com (M.
Camara Paslar)

\section{Abstract}\label{abstract}

Engine self-play allows chess openings to be compared under controlled
conditions, without the selection effects of human game databases. When
openings differ only slightly, however, rankings built from a few
hundred games can be unstable. In Study 1, strength-limited Stockfish
15.1 played 17,400 games: ten forced White first moves and ten forced
Black replies to 1.e4, each at three strength settings. The leading
opening changed between settings, twice by a single game. There was no
statistically reliable evidence, though, that differences among openings
depended on strength (White \emph{p} = .311, Black \emph{p} = .843). The
design had limited sensitivity to isolated strength-dependent effects: a
12-point effect was detected by the omnibus test in only about 39\% of
simulated studies. A Connect Four positive control detected an observed
22.5-point center-column advantage, demonstrating sensitivity to large
opening differences. With 290 games per opening, the observed leader was
the true best of ten openings in only about 44\% of simulated studies
assuming a between-opening standard deviation of three percentage
points. Follow-up studies with complete game records showed that the
settings were ordered in direct matches but did not form a consistent
Elo scale, that win and draw rates differed across hardware and
implementation configurations, and that an apparent 1.d3 weakness was
not confirmed in new games. With Stockfish 19 and a fixed node budget,
survey-selected replies showed no statistically reliable advantage over
free engine choice. Opening leaderboards from a few hundred engine games
can overstate certainty.

\textbf{Keywords:} chess openings, computer chess, engine self-play,
Stockfish, sampling error, ranking uncertainty, interaction effects,
positive control, statistical power

\section{}\label{section}

\section{1. Introduction}\label{introduction}

An opening sets the position from which the rest of a chess game
develops. Players and coaches often assume that an
opening\textquotesingle s practical value depends on who plays it.
Sharp, theory-heavy systems may reward preparation and calculation,
while quieter ones may forgive more error. This suggests a testable
question: does the relative performance of an opening change with
playing strength? One possible explanation is complexity. Openings that
gain relative to others at higher strength might demand more precise
play. Testing that would require a reliable pattern of strength
dependence and a measure of complexity defined without reference to
outcomes (Section 7).

Human game databases cannot answer this cleanly, because players choose
openings for reasons of experience, preparation, style, and reputation.
Strength-limited engine self-play gives more control. The same forced
first move can be repeated against an opponent at the same nominal
setting, without either side choosing or preparing. The cost is that a
strength-limited engine\textquotesingle s errors are not human errors,
and its nominal settings are not human ratings. This paper therefore
studies controlled engine behavior. It does not say which openings are
best for people.

Study 1 analyzed 17,400 Stockfish 15.1 self-play games from 29 valid
runs. At each of three nominal settings, the engine played ten forced
White first moves and ten forced Black replies to 1.e4, with 290 games
per opening-by-setting cell. The leading opening changed with strength.
For White, 1.e4 led at the lowest setting, 1.d4 at the middle, and 1.Nf3
at the highest. For Black, 1...e5 led at the lowest and 1...c5 at the
two higher settings. Two of these leads came down to a single game.
Taken at face value, the changes suggest that the best opening depends
on strength.

A change in the leader is not evidence that the opening effects changed.
When several openings perform about equally and each rests on a few
hundred games, chance alone can reorder them. We therefore ask two
questions: do differences among openings change systematically with
strength (an opening-by-strength interaction), and how many games are
needed to rank close alternatives reliably? We call a ranking reliable
if the same ordering would reappear in new games.

A null result is informative only if the analysis could have detected a
real effect. We therefore use a Connect Four positive control to check
detection of a large opening main effect, and sensitivity simulations to
examine power for smaller effects, including chess opening-by-strength
interactions (Sections 4.2 and 4.4). Replication and validation studies
with complete game records address the limits of Study 1.

In brief, the results are as follows. No statistically reliable
opening-by-strength interaction was found. Rankings from a few hundred
games are unstable, and the design detects only fairly large
single-opening effects. The 1.d3 deficit was not confirmed in new games.
Survey-selected replies showed no statistically reliable advantage over
free engine choice. The paper also makes four contributions: it
reconciles and checks the Study 1 summaries, tests interactions and
quantifies rank uncertainty, checks the analysis pipeline and extends
the work to a newer engine, and proposes a protocol for future
preregistration.

\section{2. Background}\label{background}

Chess research shows that expertise changes how players perceive
positions, not only how far they calculate, as de Groot (1965), Chase
and Simon (1973), and Gobet and Simon (1996) found. Error also depends
jointly on skill and position difficulty (Anderson, Kleinberg, \&
Mullainathan, 2017). These findings make a strength-dependent opening
effect plausible, but they concern human players.
Stockfish\textquotesingle s UCI\_LimitStrength option sets a target
UCI\_Elo value (Stockfish developers, 2022). At reduced strength, it
evaluates several candidate moves at shallow depth and applies a random
bias that can favor a weaker move (Stockfish developers, 2026a). This
error process differs from human error (McIlroy-Young, Sen, Kleinberg,
\& Anderson, 2020), and engine strength also depends on search depth and
hardware (Ferreira, 2013). The nominal values in this study are
therefore labels for ordered settings, not human ratings or portable
measures of strength.

Two outcome measures are used. Win percentage counts a draw as a
non-win. Expected score gives one point for a win, half a point for a
draw, and zero for a loss. If draws become more common at higher
strength, win percentages can fall even when the balance between the
colors barely changes. Connect Four provides a positive control for
detecting a large opening main effect. The game is solved: under perfect
play, the first player wins only by opening in the center column (Allis,
1988). The effect size under the study bot is estimated from its games.

Finally, Study 1 was examined before its hypotheses and analyses were
fully settled, so its inferential findings are exploratory.
Preregistration fixes hypotheses before outcomes are seen (Nosek,
Ebersole, DeHaven, \& Mellor, 2018), and flexible analysis can produce
apparently meaningful findings (Gelman \& Loken, 2014, and Kerr, 1998).
Study 4 tests the exploratory 1.d3 finding on games played after the
hypothesis was fixed. Studies 5 and 6 examine related opponent-strength
questions using a newer engine and a fixed node budget. Section 7 sets
out a protocol to be registered before new confirmatory games are
played.

\section{3. Methods}\label{methods}

\subsection{3.1 Study overview}\label{study-overview}

The research program consists of six chess studies, a Connect Four
positive control, and a separate robustness run. Study 1 provides the
original strength-limited self-play dataset. The positive control tests
whether the analysis detects a known large opening effect. Studies 2--4
address three limitations of Study 1: the lack of game records,
uncertainty about how far apart the strength settings really were, and
the exploratory status of the 1.d3 finding. Studies 5 and 6 instead hold
Black at one nominal strength while varying White\textquotesingle s
strength; they use a newer engine, a fixed node budget, independent
evaluation, and forced multi-move variations. Table 1 summarizes the
designs.

\textbf{Table 1.} Overview of the completed studies.

\begin{longtable}[]{@{}
  >{\raggedright\arraybackslash}p{(\columnwidth - 6\tabcolsep) * \real{0.1299}}
  >{\raggedright\arraybackslash}p{(\columnwidth - 6\tabcolsep) * \real{0.4500}}
  >{\raggedright\arraybackslash}p{(\columnwidth - 6\tabcolsep) * \real{0.1599}}
  >{\raggedright\arraybackslash}p{(\columnwidth - 6\tabcolsep) * \real{0.2601}}@{}}
\toprule\noalign{}
\begin{minipage}[b]{\linewidth}\raggedright
\textbf{Study}
\end{minipage} & \begin{minipage}[b]{\linewidth}\raggedright
\textbf{Purpose and comparison}
\end{minipage} & \begin{minipage}[b]{\linewidth}\raggedright
\textbf{Scheduled games}
\end{minipage} & \begin{minipage}[b]{\linewidth}\raggedright
\textbf{Outcome and status}
\end{minipage} \\
\begin{minipage}[b]{\linewidth}\raggedright
Study 1
\end{minipage} & \begin{minipage}[b]{\linewidth}\raggedright
Ten White first moves and ten Black replies at three Stockfish 15.1
strength settings, with both sides at the same setting
\end{minipage} & \begin{minipage}[b]{\linewidth}\raggedright
18,000
\end{minipage} & \begin{minipage}[b]{\linewidth}\raggedright
Designated-color win (exploratory)
\end{minipage} \\
\begin{minipage}[b]{\linewidth}\raggedright
Study 2
\end{minipage} & \begin{minipage}[b]{\linewidth}\raggedright
Study 1 design repeated with complete game records and 30 games per cell
\end{minipage} & \begin{minipage}[b]{\linewidth}\raggedright
1,800
\end{minipage} & \begin{minipage}[b]{\linewidth}\raggedright
Wins and expected score (replication and robustness)
\end{minipage} \\
\begin{minipage}[b]{\linewidth}\raggedright
Study 3
\end{minipage} & \begin{minipage}[b]{\linewidth}\raggedright
Direct matches among the three settings, with colors alternated
\end{minipage} & \begin{minipage}[b]{\linewidth}\raggedright
800
\end{minipage} & \begin{minipage}[b]{\linewidth}\raggedright
Stronger setting\textquotesingle s expected score
\end{minipage} \\
\begin{minipage}[b]{\linewidth}\raggedright
Study 4
\end{minipage} & \begin{minipage}[b]{\linewidth}\raggedright
New games testing whether 1.d3 scores below the other White moves (the
test also uses the White-opening games from Study 2)
\end{minipage} & \begin{minipage}[b]{\linewidth}\raggedright
1,800 additional
\end{minipage} & \begin{minipage}[b]{\linewidth}\raggedright
White win and expected score (hypothesis fixed in advance but not
publicly registered)
\end{minipage} \\
\begin{minipage}[b]{\linewidth}\raggedright
Study 5
\end{minipage} & \begin{minipage}[b]{\linewidth}\raggedright
Reply survey, followed by independent evaluation of the selected
replies, a fixed book, and free Stockfish
\end{minipage} & \begin{minipage}[b]{\linewidth}\raggedright
6,960 survey and 5,220 evaluation
\end{minipage} & \begin{minipage}[b]{\linewidth}\raggedright
Black expected score (separate selection and evaluation data)
\end{minipage} \\
\begin{minipage}[b]{\linewidth}\raggedright
Study 6
\end{minipage} & \begin{minipage}[b]{\linewidth}\raggedright
Eighteen forced multi-move variations against two White strength
settings
\end{minipage} & \begin{minipage}[b]{\linewidth}\raggedright
10,440
\end{minipage} & \begin{minipage}[b]{\linewidth}\raggedright
Black expected score (exploratory)
\end{minipage} \\
\begin{minipage}[b]{\linewidth}\raggedright
Connect Four
\end{minipage} & \begin{minipage}[b]{\linewidth}\raggedright
Seven forced opening columns, with the weakest bot on both sides
\end{minipage} & \begin{minipage}[b]{\linewidth}\raggedright
2,100
\end{minipage} & \begin{minipage}[b]{\linewidth}\raggedright
First-player win (positive control)
\end{minipage} \\
\midrule\noalign{}
\endhead
\bottomrule\noalign{}
\endlastfoot
\end{longtable}

\emph{Note. Study 1 analyzed 17,400 of 18,000 scheduled chess games, and
Connect Four analyzed 2,030 of 2,100, because one run\textquotesingle s
output was lost. A separate Studies 2--4 robustness run used a second
computing environment and different randomized game orders; it is not a
primary study.}

\subsection{3.2 Study 1: design and data}\label{study-1-design-and-data}

\subsubsection{Engine configuration and
treatments}\label{engine-configuration-and-treatments}

Study 1 games were generated by the simulation module of a web-based
game platform, using the configuration in source-code commit 5872d85.
The server ran Stockfish 15.1 with UCI\_LimitStrength enabled, one
thread, a 1 MB hash table, and a fixed search time of 10 milliseconds
per move. Nominal UCI\_Elo values of 1320, 2000, and 2500 were
requested. Because 1320 is below the minimum allowed by this release,
the lowest setting actually ran at the 1350 floor. The values are
treated as labels for three settings, not as calibrated Elo ratings.
Games ran on a shared free-tier processor, so the number of positions
searched within the time limit depended on the hardware.

Each run contained 600 chess games, 200 at each setting (Supplementary
Figure S1). At each setting, 100 White-opening games forced one of ten
White first moves, ten games per move. A separate set of 100 Black-reply
games forced 1.e4 for White and then one of ten Black replies, again ten
games per reply. Apart from the forced move, both sides used the same
strength setting. The full list of treatments and simulator labels
appears in Supplementary Table S2. The White treatments therefore fixed
one move, while the Black treatments fixed the first two plies
(half-moves). These are initial commitments, not complete named opening
variations.

The engine received the current position as a FEN string on every move,
rather than the full move history. Games ended by checkmate, stalemate,
insufficient material, fivefold repetition, or the 75-move rule. Games
that reached 300 plies were stopped and scored as draws. Supplementary
Table S1 gives the full configuration, including concurrency, seed
handling, library behavior, and source-code locations.

\subsubsection{Runs, exclusions, and data
recovery}\label{runs-exclusions-and-data-recovery}

Thirty runs were planned. The output of one completed run (Run 7) was
lost before it could be recorded, so no data from it were analyzed. This
left 29 valid runs, 17,400 games, and 290 games in each of the 60
opening-by-setting cells. No values were imputed and no other runs were
excluded.

Study 1 did not save game records. Cell counts were reconstructed from
archived screenshots of each run\textquotesingle s summary and bar
charts and summed across runs. Setting-level summaries were transcribed
independently, giving 87 run-by-setting records. Run 12 at the 2000
setting lacked its summary panel, so its draw rate is missing. The six
setting-by-color totals matched exactly. This rules out large systematic
errors, but not errors that cancel out within a run and setting.

\subsection{3.3 Study 1: outcomes and
analyses}\label{study-1-outcomes-and-analyses}

\subsubsection{Primary outcome and central
estimand}\label{primary-outcome-and-central-estimand}

The primary outcome for each cell was the designated-color win
probability. This is the probability that White won after a forced White
first move, or that Black won after a forced Black reply. Draws stayed
in the denominator because Study 1 lacked per-opening draw counts. The
central question (the estimand) was whether the differences among
openings changed across settings, which is the opening-by-strength
interaction. This is different from an overall change in win rate that
affects most openings together. For each opening, its relative change
was also summarized by a \emph{difference-in-differences} contrast: its
change from the lowest to the highest setting, minus the average change
of the other nine openings of the same color.

Separate binomial logistic models were fitted for White and Black. An
additive model with opening and strength was compared, by
likelihood-ratio test, with a model that also included their
interaction. Strength was treated as categorical. As a focused
sensitivity analysis, a second model allowed each
opening\textquotesingle s effect to change linearly across the three
ordered settings. Opening-specific contrasts used normal approximations,
with Holm adjustment across the ten openings of each color. The less
conservative Benjamini--Hochberg adjustment (Benjamini \& Hochberg,
1995) was also reported as a sensitivity analysis.

\subsubsection{Secondary and diagnostic
analyses}\label{secondary-and-diagnostic-analyses}

Cell estimates use 95\% Wilson intervals (Wilson, 1927, and Agresti,
2013). Within each color and setting, heterogeneity among the ten
openings was tested with a 2 × 10 Pearson chi-square test. Because the
range between the highest and lowest win rates has no simple reference
distribution, its null distribution was simulated from 100,000 samples
of ten equal-probability binomial cells. Holm\textquotesingle s (1979)
procedure adjusts the six within-setting tests. Ranking reliability is
the between-opening variance divided by the total variance, so values
near 0 mean the ordering is mostly noise.

The 1.d3 contrast was tested with a Mantel--Haenszel common odds ratio
stratified by setting (Mantel \& Haenszel, 1959), with a test of effect
homogeneity. Because 1.d3 was selected as the weakest move after
inspection, a Bonferroni adjustment for choosing one of ten candidates
was also applied.

Rank uncertainty was evaluated with 20,000 parametric bootstrap samples
drawn from the observed cell proportions (Efron \& Tibshirani, 1993).
Because this bootstrap is centered on the observed rates, it favors the
observed leaders, so its probabilities are optimistic. An
empirical-Bayes shrinkage analysis gave a less optimistic alternative.

Run-level summaries provided independent replications of each
setting-level rate. The displayed draw rate combined both sets, so
White\textquotesingle s first-move advantage could not be recovered
exactly. If \emph{W} is White\textquotesingle s win rate in the White
set, \emph{B} Black\textquotesingle s win rate in the Black set, and
\emph{D} the combined draw rate, the advantage lies between \emph{W} −
\emph{B} − \emph{D} and \emph{W} − \emph{B} + \emph{D} (Appendix A).

Power was estimated by normal approximation and by simulation (Section
4.2). Analyses used Python 3.12 with NumPy, pandas, SciPy (Virtanen et
al., 2020), and statsmodels (Seabold \& Perktold, 2010), with fixed
random seeds.

\subsection{3.4 Positive control and follow-up
studies}\label{positive-control-and-follow-up-studies}

\subsubsection{Connect Four positive
control}\label{connect-four-positive-control}

Each run also included 70 Connect Four games, ten per forced first-move
column, with the weakest bot on both sides. This gives 2,030 games, or
290 per column. They were analyzed with chess main-effect procedures:
Wilson intervals, a chi-square homogeneity test, a
center-versus-other-columns contrast, and ranking reliability. This
control does not test opening-by-strength interactions.

\subsubsection{Study 2: replication with complete
records}\label{study-2-replication-with-complete-records}

An independent Python program using python-chess 1.11 (Fiekas, n.d.)
reproduced the platform\textquotesingle s game loop and recorded every
move list, result, termination reason, and game length. Study 2 used
Stockfish 15.1 with the same requested settings, 10-millisecond move
limit, FEN-only input, one thread, 1 MB hash, and 300-ply cap. Thirty
games were played in each of the 60 Study 1 cells, for 1,800 games in
all, in randomized order. The study ran on a dedicated modern laptop,
which would be expected to search more positions in 10 milliseconds than
the shared Study 1 server. Study 2 therefore tests whether the relative
opening conclusions hold under the same software settings, not whether
absolute outcome rates reproduce across hardware. Because Study 2 also
used an independently written game loop, any change in absolute rates
could be due to either the hardware or the implementation.

\subsubsection{Study 3: actual strength
gaps}\label{study-3-actual-strength-gaps}

Direct matches tested whether the settings really played at different
strengths. Colors alternated in 200 games between the lowest and middle
settings, 200 games between the middle and highest settings, and 400
games between the lowest and highest settings. The stronger
setting\textquotesingle s score (wins plus half of draws) was converted
to an Elo difference using the logistic formula, with intervals based on
the score\textquotesingle s standard error. The widest nominal pairing
received more games because an earlier run had produced an unexpectedly
small estimated gap. This allocation therefore depended on earlier data.

\subsubsection{Study 4: prospective test of the 1.d3
deficit}\label{study-4-prospective-test-of-the-1.d3-deficit}

Before any follow-up games were played, an earlier version of this
manuscript fixed the hypothesis that 1.d3 would score below the other
nine White moves. It also fixed the test (a one-sided Mantel--Haenszel
test stratified by setting) and the significance level (α = .05). The
hypothesis was not entered in a public registry. Study 4 added 60 games
per White move per setting, or 1,800 games. The test combined these
games with the 900 White-opening games from Study 2, giving 90 new games
per move per setting. This gave about 70\% power if the true effect
matched the exploratory estimate. Because that estimate was the most
extreme of ten candidates, the true deficit is probably smaller, and
power for a smaller deficit is lower. Monte Carlo simulation of the
prespecified test, at a 40\% win rate for the comparison moves, gave
about 36\% power for a four-point gap and 24\% for a three-point gap.
The test used the original 10-millisecond configuration rather than the
more demanding configuration proposed in Section 7.

\subsubsection{Independent robustness
run}\label{independent-robustness-run}

A separate run of the same program on a single-core cloud processor,
with different randomized orders, is reported in Section 4.4 and
excluded from the primary analyses. It reveals instability from
execution or hardware but cannot rule out errors shared by both runs.

\subsection{3.5 Studies 5 and 6: fixed-compute
extensions}\label{studies-5-and-6-fixed-compute-extensions}

Studies 5 and 6 used Stockfish 19 (Stockfish developers, 2026b) with
UCI\_LimitStrength, one thread, a 16 MB hash table, and a fixed budget
of 3,000 nodes per move. A node limit makes the amount of search
independent of processor speed. Standard draw rules, including threefold
repetition, were applied. Games that reached 400 plies without a result
were censored (set aside) rather than scored as draws. Black always
played at nominal 2000, and White played at either 1800 or 2000. These
studies therefore ask whether Black\textquotesingle s best opening
choice depends on the opponent\textquotesingle s strength. Cells were
scheduled in 29 randomized blocks of ten games, and differences between
blocks were used for inference. The absolute rates in these studies
cannot be compared directly with those of Studies 1--4.

\subsubsection{Study 5: selection and independent
evaluation}\label{study-5-selection-and-independent-evaluation}

White\textquotesingle s first move was forced to 1.e4, 1.d4, or 1.c4. In
a survey phase, four prespecified Black replies were compared for each
White move at both White settings: 24 cells of 290 games, or 6,960
games. For each of the six combinations of White move and setting, a
fixed rule selected the reply with the highest smoothed expected score.
The resulting set of choices (the "learned policy") was then frozen. An
independent evaluation crossed the three White moves and two White
settings with three ways of choosing Black\textquotesingle s reply: the
learned policy, a prespecified fixed book, and letting Stockfish choose
freely. This produced 18 new cells and 5,220 games. The two prespecified
primary contrasts compared the learned policy with the free engine at
each White setting, weighting the three first moves equally and applying
Holm adjustment across the two tests.

\subsubsection{Study 6: forced multi-move
variations}\label{study-6-forced-multi-move-variations}

Study 6 forced 18 multi-move sequences from six opening families: four
Sicilian, four Ruy Lopez, four Queen\textquotesingle s Gambit, two
French, two Caro-Kann, and two English variations. Each variation was
tested against White at 1800 and at 2000, producing 36 cells of 290
games and 10,440 scheduled games. The outcome was
Black\textquotesingle s expected score among completed games. Pairs
within the same family were compared with paired \emph{t} tests on the
29 block differences, with Holm adjustment across 42 strength-specific
comparisons. The 21 variation-by-strength interaction tests were
adjusted separately. These comparisons were exploratory. Full sequences,
schedules, candidate policies, and configuration fingerprints are
provided in the supplementary material.

\subsection{3.6 Reproducibility}\label{reproducibility}

The Zenodo archive (Flick et al., 2026,
https://doi.org/10.5281/zenodo.23357406) contains reconstructed Study 1
summaries, game records for the follow-up studies, protocols, and
analysis code. The Declarations describe its contents.

\section{4. Results: Opening Performance, Ranking Uncertainty, and
Validation}\label{results-opening-performance-ranking-uncertainty-and-validation}

Sections 4.1--4.3 cover Study 1 outcomes, strength interactions, and
ranking uncertainty. Section 4.4 reports validation and replication;
Section 4.5 presents fixed-compute extensions involving unequal-strength
opponents.

\subsection{4.1 Data integrity and setting-level
outcomes}\label{data-integrity-and-setting-level-outcomes}

\subsubsection{Data integrity}\label{data-integrity}

Study 1 contained 290 games in each of the 60 opening-by-setting cells.
White won 1,478, 1,316, and 1,230 of the 2,900 White-opening games at
the 1320, 2000, and 2500 settings. Black won 1,281, 1,140, and 1,128 of
the corresponding Black-reply games. These six totals matched the
independently transcribed run summaries exactly, and no duplicate run
records were found. This agreement makes a large systematic
transcription error unlikely.

\subsubsection{Wins, draws, and first-move
advantage}\label{wins-draws-and-first-move-advantage}

As the nominal setting increased, designated-color win rates fell and
draws became much more common (Table 2 and Figure 1).
White\textquotesingle s win rate in the White-opening set fell from
51.0\% at 1320 to 45.4\% at 2000 and 42.4\% at 2500. The decline from
1320 to 2500 was consistent across runs, \emph{t}(28) = 5.49, \emph{p}
\textless{} .001. Black\textquotesingle s win rate in the Black-reply
set fell from 44.2\% to 39.3\% and 38.9\%. Over the same settings, the
pooled draw rate rose from 1.1\% to 12.8\% and 16.2\% (Kruskal--Wallis
\emph{p} \textless{} .001, and \emph{p} \textless{} .001 for 2000 versus
2500).

\textbf{Table 2. Study 1 outcomes by nominal strength setting (1320
denotes the effective 1350 floor).}

\begin{longtable}[]{@{}
  >{\raggedright\arraybackslash}p{(\columnwidth - 8\tabcolsep) * \real{0.1200}}
  >{\raggedright\arraybackslash}p{(\columnwidth - 8\tabcolsep) * \real{0.2200}}
  >{\raggedright\arraybackslash}p{(\columnwidth - 8\tabcolsep) * \real{0.2200}}
  >{\raggedright\arraybackslash}p{(\columnwidth - 8\tabcolsep) * \real{0.2200}}
  >{\raggedright\arraybackslash}p{(\columnwidth - 8\tabcolsep) * \real{0.2201}}@{}}
\toprule\noalign{}
\begin{minipage}[b]{\linewidth}\raggedright
\textbf{Setting}
\end{minipage} & \begin{minipage}[b]{\linewidth}\raggedright
\textbf{White wins, White set}
\end{minipage} & \begin{minipage}[b]{\linewidth}\raggedright
\textbf{Black wins, Black set}
\end{minipage} & \begin{minipage}[b]{\linewidth}\raggedright
\textbf{Draws}
\end{minipage} & \begin{minipage}[b]{\linewidth}\raggedright
\textbf{Implied White expected score\textsuperscript{a}}
\end{minipage} \\
\begin{minipage}[b]{\linewidth}\raggedright
1320
\end{minipage} & \begin{minipage}[b]{\linewidth}\raggedright
51.0\% {[}48.9, 53.1{]}
\end{minipage} & \begin{minipage}[b]{\linewidth}\raggedright
44.2\% {[}42.1, 46.2{]}
\end{minipage} & \begin{minipage}[b]{\linewidth}\raggedright
1.1\% {[}0.8, 1.4{]}
\end{minipage} & \begin{minipage}[b]{\linewidth}\raggedright
53.4\%
\end{minipage} \\
\begin{minipage}[b]{\linewidth}\raggedright
2000
\end{minipage} & \begin{minipage}[b]{\linewidth}\raggedright
45.4\% {[}44.0, 46.7{]}
\end{minipage} & \begin{minipage}[b]{\linewidth}\raggedright
39.3\% {[}37.7, 40.9{]}
\end{minipage} & \begin{minipage}[b]{\linewidth}\raggedright
12.8\% {[}11.9, 13.7{]}
\end{minipage} & \begin{minipage}[b]{\linewidth}\raggedright
53.0\%
\end{minipage} \\
\begin{minipage}[b]{\linewidth}\raggedright
2500
\end{minipage} & \begin{minipage}[b]{\linewidth}\raggedright
42.4\% {[}40.2, 44.6{]}
\end{minipage} & \begin{minipage}[b]{\linewidth}\raggedright
38.9\% {[}37.2, 40.6{]}
\end{minipage} & \begin{minipage}[b]{\linewidth}\raggedright
16.2\% {[}15.2, 17.2{]}
\end{minipage} & \begin{minipage}[b]{\linewidth}\raggedright
51.8\%
\end{minipage} \\
\midrule\noalign{}
\endhead
\bottomrule\noalign{}
\endlastfoot
\end{longtable}

\emph{Note.} Bracketed values are 95\% t intervals across runs. The 2000
draw estimate is based on 28 runs because one run summary was
unavailable. \textsuperscript{a}Assumes equal draw rates in the
White-opening and Black-reply sets at each setting.

\includegraphics[width=6in,height=2.34375in]{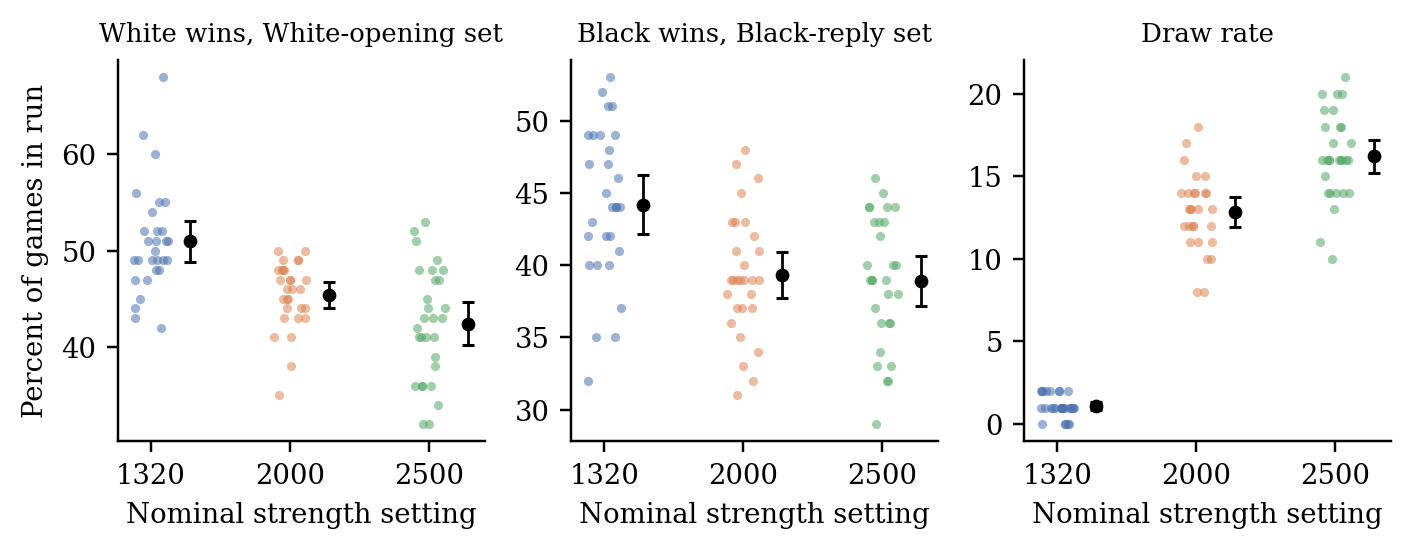}

\textbf{Figure 1.} Run-level outcome rates by setting. Each colored
point is one run. Black points and bars show the mean and 95\% t
interval across runs. Win rates are based on 100 games per run and
setting, and draw rates use all 200 games at a setting.

The rise in draws explains most of the drop in designated-color wins. If
the White-opening and Black-reply sets are assumed to have equal draw
rates within a setting, White\textquotesingle s expected score changes
much less than its win rate: 53.4\%, 53.0\%, and 51.8\%. The lower win
rates at higher settings should therefore not be read as a matching
decline in the quality of White\textquotesingle s positions. Study 2
later shows that this absolute draw pattern did not reproduce on
different hardware with an independently written game loop.

At 1320, where draws were rare, the available summaries bound
White\textquotesingle s first-move advantage between 5.7 and 7.9
percentage points, and its expected score between 52.8\% and 53.9\%. At
2000 and 2500, the bounds include zero, because Study 1 did not record
draw counts separately for the two sets. Assuming equal draw rates gives
first-move-advantage estimates of 6.1 and 3.5 points (Supplementary
Table S9). The complete Study 2 records give direct, though less
precise, estimates (Section 4.4).

\subsubsection{Setting separation and run-level
dispersion}\label{setting-separation-and-run-level-dispersion}

Run-to-run dispersion ratios ranged from 0.50 to 1.41, and none of the
nine overdispersion tests reached .05 (Supplementary Table S10). The
largest, for White wins at 2500, gave \emph{p = .075.} These checks
support the binomial models at the setting level, but they cannot detect
dependence confined to one opening. Study 3 is the stronger test of
strength ordering.

\subsection{4.2 Opening differences and strength
dependence}\label{opening-differences-and-strength-dependence}

\subsubsection{Differences within
settings}\label{differences-within-settings}

Uncertainty at the level of individual openings was large (Supplementary
Figures S2 and S3). With 290 games per cell, a typical 95\% Wilson
interval spans about 11 percentage points. The observed ranges among
White moves were 12.1, 13.4, and 13.8 points at the three settings,
compared with 8.3, 8.6, and 10.3 points among Black replies. In
simulations where all ten openings had the same true win probability,
the typical observed range was about 9 points, and a range near 13
points occurred about 5\% of the time (Supplementary Figure S4). The
White spreads are therefore somewhat larger than chance alone would
usually produce, while the Black spreads are consistent with chance.

Separate homogeneity tests pointed in the same direction but were not
conclusive (Supplementary Table S4). The White tests gave \emph{p} =
.191, .077, and .029 at 1320, 2000, and 2500. The Black tests gave
\emph{p} = .590, .287, and .454. None remained significant after Holm
adjustment across the six tests (smallest adjusted \emph{p} = .174).
Estimated ranking reliability for White rose across the settings from
.28 to .42 and .52, but for Black it was .00, .17, and .00. Even where
some White moves appear to differ, sampling error still accounts for
much of the observed ordering.

\subsubsection{Central interaction test}\label{central-interaction-test}

The main analysis asked whether the differences among openings changed
across strength settings. Strength had a clear main effect on
designated-color wins for both colors (Table 3). After adjusting for
strength, the White moves also differed overall, LR χ²(9) = 26.19,
\emph{p} = .002, but the Black replies did not, LR χ²(9) = 15.01,
\emph{p} = .091. Much of the White opening effect comes from 1.d3 (see
below). These main effects describe average differences among openings.
They do not show that the openings\textquotesingle{} relative
performance changes with strength.

\textbf{Table 3.} Likelihood-ratio tests for opening and strength
effects in Study 1.

\begin{longtable}[]{@{}
  >{\raggedright\arraybackslash}p{(\columnwidth - 8\tabcolsep) * \real{0.1200}}
  >{\raggedright\arraybackslash}p{(\columnwidth - 8\tabcolsep) * \real{0.4599}}
  >{\raggedright\arraybackslash}p{(\columnwidth - 8\tabcolsep) * \real{0.1400}}
  >{\raggedright\arraybackslash}p{(\columnwidth - 8\tabcolsep) * \real{0.1000}}
  >{\raggedright\arraybackslash}p{(\columnwidth - 8\tabcolsep) * \real{0.1801}}@{}}
\toprule\noalign{}
\begin{minipage}[b]{\linewidth}\raggedright
\textbf{Color}
\end{minipage} & \begin{minipage}[b]{\linewidth}\raggedright
\textbf{Term tested}
\end{minipage} & \begin{minipage}[b]{\linewidth}\raggedright
\textbf{LR χ²}
\end{minipage} & \begin{minipage}[b]{\linewidth}\raggedright
\textbf{df}
\end{minipage} & \begin{minipage}[b]{\linewidth}\raggedright
\textbf{p}
\end{minipage} \\
\begin{minipage}[b]{\linewidth}\raggedright
White
\end{minipage} & \begin{minipage}[b]{\linewidth}\raggedright
Strength main effect
\end{minipage} & \begin{minipage}[b]{\linewidth}\raggedright
43.99
\end{minipage} & \begin{minipage}[b]{\linewidth}\raggedright
2
\end{minipage} & \begin{minipage}[b]{\linewidth}\raggedright
\textless{} .001
\end{minipage} \\
\begin{minipage}[b]{\linewidth}\raggedright
White
\end{minipage} & \begin{minipage}[b]{\linewidth}\raggedright
Opening main effect
\end{minipage} & \begin{minipage}[b]{\linewidth}\raggedright
26.19
\end{minipage} & \begin{minipage}[b]{\linewidth}\raggedright
9
\end{minipage} & \begin{minipage}[b]{\linewidth}\raggedright
.002
\end{minipage} \\
\begin{minipage}[b]{\linewidth}\raggedright
White
\end{minipage} & \begin{minipage}[b]{\linewidth}\raggedright
Opening × strength (general)
\end{minipage} & \begin{minipage}[b]{\linewidth}\raggedright
20.40
\end{minipage} & \begin{minipage}[b]{\linewidth}\raggedright
18
\end{minipage} & \begin{minipage}[b]{\linewidth}\raggedright
.311
\end{minipage} \\
\begin{minipage}[b]{\linewidth}\raggedright
White
\end{minipage} & \begin{minipage}[b]{\linewidth}\raggedright
Opening × linear strength trend
\end{minipage} & \begin{minipage}[b]{\linewidth}\raggedright
6.89
\end{minipage} & \begin{minipage}[b]{\linewidth}\raggedright
9
\end{minipage} & \begin{minipage}[b]{\linewidth}\raggedright
.649
\end{minipage} \\
\begin{minipage}[b]{\linewidth}\raggedright
Black
\end{minipage} & \begin{minipage}[b]{\linewidth}\raggedright
Strength main effect
\end{minipage} & \begin{minipage}[b]{\linewidth}\raggedright
20.60
\end{minipage} & \begin{minipage}[b]{\linewidth}\raggedright
2
\end{minipage} & \begin{minipage}[b]{\linewidth}\raggedright
\textless{} .001
\end{minipage} \\
\begin{minipage}[b]{\linewidth}\raggedright
Black
\end{minipage} & \begin{minipage}[b]{\linewidth}\raggedright
Opening main effect
\end{minipage} & \begin{minipage}[b]{\linewidth}\raggedright
15.01
\end{minipage} & \begin{minipage}[b]{\linewidth}\raggedright
9
\end{minipage} & \begin{minipage}[b]{\linewidth}\raggedright
.091
\end{minipage} \\
\begin{minipage}[b]{\linewidth}\raggedright
Black
\end{minipage} & \begin{minipage}[b]{\linewidth}\raggedright
Opening × strength (general)
\end{minipage} & \begin{minipage}[b]{\linewidth}\raggedright
12.09
\end{minipage} & \begin{minipage}[b]{\linewidth}\raggedright
18
\end{minipage} & \begin{minipage}[b]{\linewidth}\raggedright
.843
\end{minipage} \\
\begin{minipage}[b]{\linewidth}\raggedright
Black
\end{minipage} & \begin{minipage}[b]{\linewidth}\raggedright
Opening × linear strength trend
\end{minipage} & \begin{minipage}[b]{\linewidth}\raggedright
7.88
\end{minipage} & \begin{minipage}[b]{\linewidth}\raggedright
9
\end{minipage} & \begin{minipage}[b]{\linewidth}\raggedright
.546
\end{minipage} \\
\midrule\noalign{}
\endhead
\bottomrule\noalign{}
\endlastfoot
\end{longtable}

\emph{Note.} Strength is tested against an intercept-only model, opening
against a strength-only model, and each interaction against the additive
model. The linear-trend test allows each opening\textquotesingle s logit
effect to change linearly across the three ordered settings.

No statistically reliable opening-by-strength interaction was found. For
White, the general interaction test gave \emph{p} = .311 and the focused
linear-trend test gave \emph{p} = .649. For Black, the corresponding
values were .843 and .546. The additive models describe the cell counts
adequately. In other words, the data are consistent with all openings
shifting together on the logit scale as the setting changes
(Supplementary Figure S5). This does not prove that all interaction
effects are zero. Realistic small differences remain unresolved with
these sample sizes.

To estimate what the omnibus tests could have detected, their power was
approximated with a normal approximation. This assumed independent
interaction effects with a common standard deviation across the 30
opening-by-setting cells of each color, and a mean win probability of
45\%. Under these assumptions, the 18-df tests would have had roughly
10\%, 35\%, and 77\% power for interaction standard deviations of 1, 2,
and 3 percentage points, and about 98\% power at 4 points. A single
opening that departs from the others is harder to detect. Its contrast
would need to be roughly 12 to 16 points to reach 80\% power. A
simulation using binomial logistic likelihood-ratio tests gave similar
power values of 11\%, 36\%, 74\%, and 93\%. For a single opening,
however, the omnibus test is weak. A simulation of one opening with a
12-point effect detects it in only about 39\% of studies (Table 4). The
approximations above describe diffuse interactions across many openings,
and they should not be read as power for one opening.

The opening-specific comparisons lead to the same conclusion (Figure 2
and Supplementary Tables S5 and S6). The only absolute change from 1320
to 2500 that survived Holm adjustment was the decline for 1.Nc3, from
51.4\% to 39.0\% (−12.4 points, 95\% CI {[}−20.5, −4.4{]}, adjusted
\emph{p} = .025). Relative to the average change of the other White
moves, however, its decline was only −4.3 points (95\% CI {[}−12.8,
4.2{]}). The largest relative changes were +6.8 points for 1.Nf3, +6.4
for 1.b3, and +7.8 for Black\textquotesingle s 1...c5, but every
interval included zero. No difference-in-differences contrast survived
Holm or Benjamini--Hochberg adjustment.

\includegraphics[width=6in,height=3.07292in]{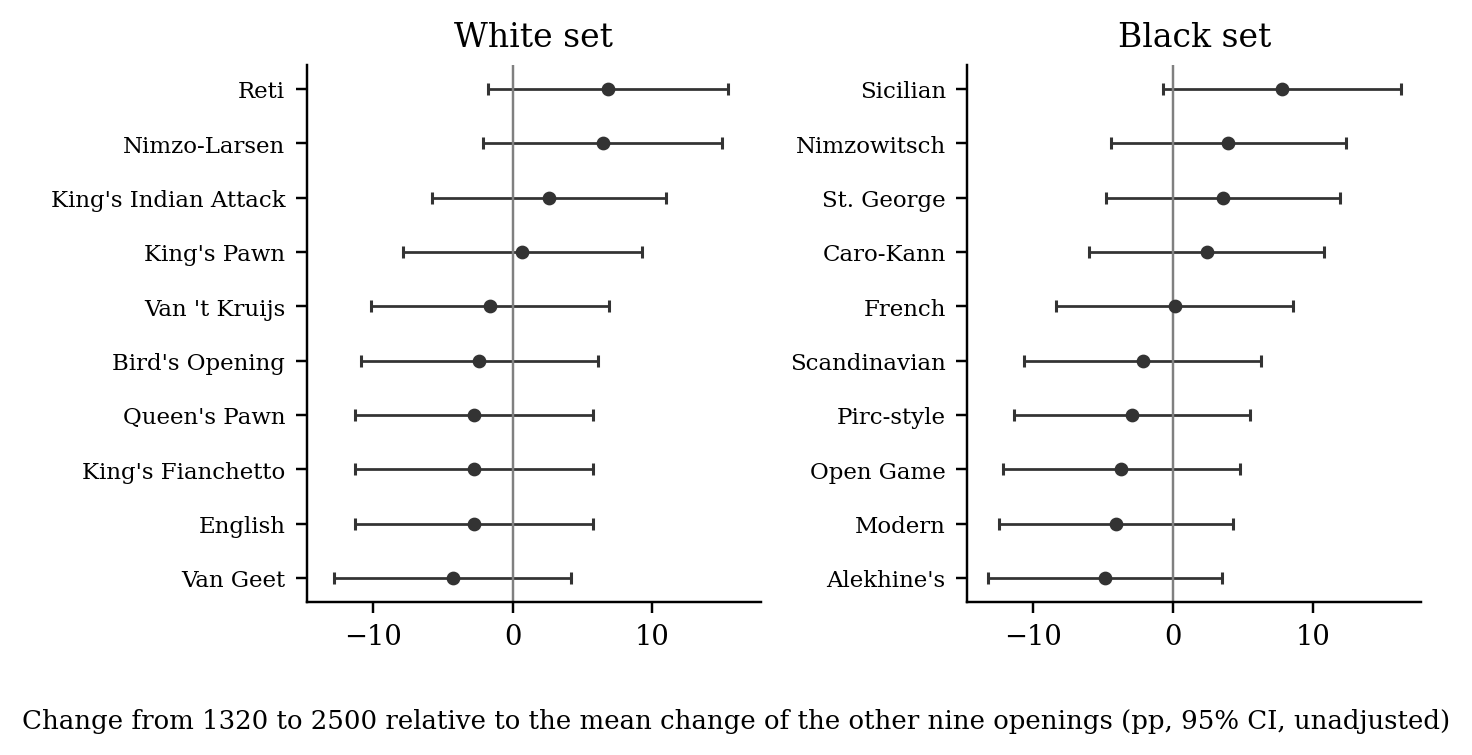}

\textbf{Figure 2.} Each opening\textquotesingle s change in win
percentage from 1320 to 2500, minus the average change of the other nine
openings of the same color. Intervals are unadjusted 95\% confidence
intervals, and all of them include zero.

\subsubsection{Exploratory 1.d3 finding}\label{exploratory-1.d3-finding}

Much of the White opening main effect came from 1.d3. It was the
lowest-scoring White move at 1320 and 2500 and the second lowest at
2000. Pooled across settings, 1.d3 won 40.0\% of 870 games, compared
with 46.9\% for the other nine moves (Supplementary Table S7). The
stratified common odds ratio was 0.75 (95\% CI {[}0.65, 0.87{]},
\emph{p} \textless{} .001), with no sign that the effect differed across
settings (\emph{p} = .523). The result stayed significant after
correcting for selecting the weakest of ten moves. Because the contrast
was identified after looking at Study 1, it is treated as exploratory
and tested prospectively in Section 4.4.

\subsection{4.3 Ranking uncertainty and required sample
size}\label{ranking-uncertainty-and-required-sample-size}

\subsubsection{Uncertainty in the observed
leaders}\label{uncertainty-in-the-observed-leaders}

Which opening came first was much less stable than the overall opening
effects. In the parametric bootstrap, the observed White leader ranked
first in 29\% of replicates at 1320, 74\% at 2000, and 68\% at 2500
(Supplementary Figure S6). Because these simulations are centered on the
observed rates, they favor the observed leaders. After empirical-Bayes
shrinkage, the corresponding probabilities fell to .21, .53, and .54.
For Black, the estimated between-opening variance was zero at 1320 and
2500, which gives every reply a .10 probability of being first. At 2000,
the observed Black leader\textquotesingle s probability was .21. Most
95\% rank intervals covered at least six of the ten positions
(Supplementary Tables S7 and S8).

Spearman rank correlations between settings ranged from .17 to .61 for
White and from .16 to .71 for Black. These modest and variable
correlations do not by themselves show that the underlying rankings
changed. When true differences between openings are small relative to
sampling error, observed rankings can shift a great deal even if the
true ordering stays the same.

\subsubsection{Games required to identify the
leader}\label{games-required-to-identify-the-leader}

Simulations showed what this uncertainty means for sample size (Figure
3). The results can be reproduced by simulating ten openings whose true
win probabilities are normally distributed, with a mean of 45\% and a
between-opening standard deviation of 2, 3, or 4 percentage points. With
the current 290 games per opening and a 3-point standard deviation, the
opening with the highest observed rate was truly the best in about 44\%
of 20,000 simulations. The figure was about 33\% with a 2-point and 52\%
with a 4-point standard deviation. Reaching an 80\% chance of
identifying the best opening required about 5,000 games per opening when
the standard deviation was 3 points, and about 10,000 when it was 2
points. Even at 20,000 games per opening, the observed leader was wrong
in 7\% to 15\% of simulations. These values depend on the stated model
and should not be treated as universal thresholds.

The between-opening standard deviations implied by the Study 1
reliability estimates (Section 4.2) were about 1.8 to 3.0 points for
White and 0 to 1.3 points for Black. The 3-point scenario is therefore
near the top of the White estimates and above the Black estimates. Where
true differences are smaller, leaderboards are even less reliable than
the 44\% figure suggests.

\includegraphics[width=6in,height=2.9375in]{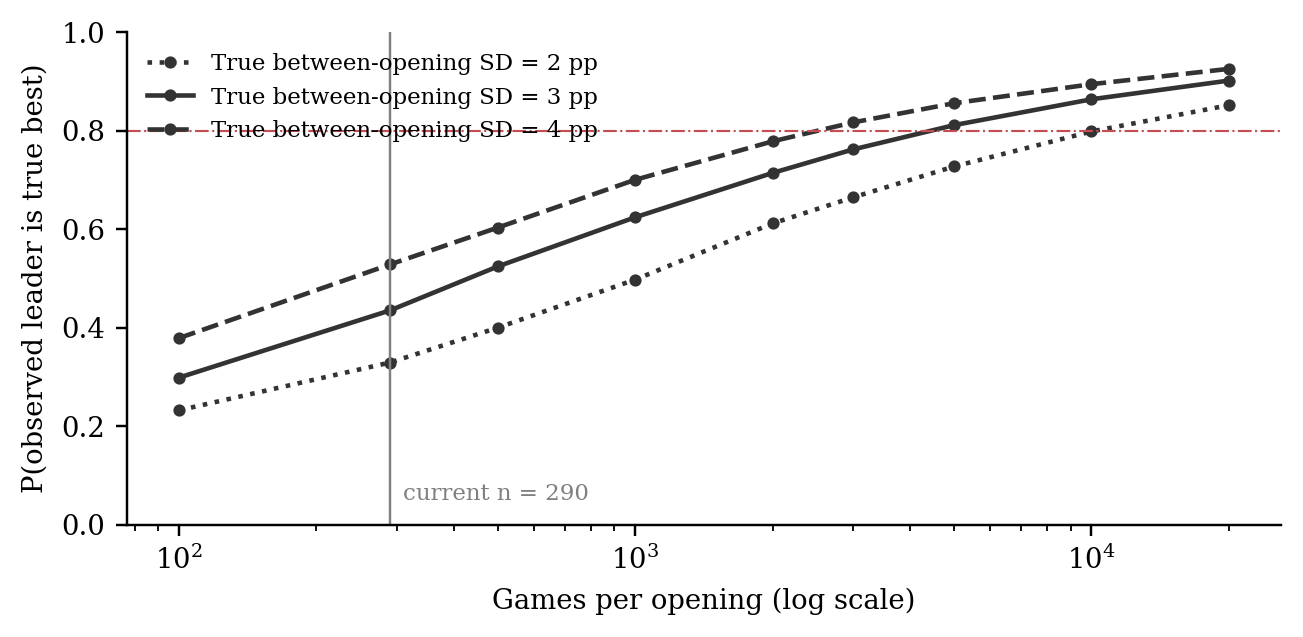}

\textbf{Figure 3.} Probability that the observed leader is the true best
of ten openings, under three assumed levels of between-opening variation
(20,000 simulated studies per point, mean win probability 45\%). The
vertical line marks 290 games per opening, and the horizontal line marks
80\% probability.

A less demanding goal is easier to reach. With 290 games per opening and
a 3-point standard deviation, the truly best opening was among the
observed top three in 78\% of simulations. The evidence therefore
supports drawing up a shortlist of plausible leaders more readily than
naming a single best opening.

\subsection{4.4 Validation and
replication}\label{validation-and-replication}

\subsubsection{Connect Four positive
control}\label{connect-four-positive-control-1}

The first player won 63.1\% of games from the center column (95\% CI
{[}57.4, 68.5{]}, or 183 of 290 games). The other six columns ranged
from 32.1\% to 45.2\%. The homogeneity test across columns was
significant, χ²(6) = 64.47, \emph{p} \textless{} .001. The center column
exceeded the mean of the other six by 22.5 points (95\% CI {[}16.5,
28.5{]}). Ranking reliability was .91, and in a parametric simulation
the center column led in every replicate (Figure 4). Because each run
contributed only ten games per column, the center column had the highest
score, counting ties, in 20 of the 29 runs. The control therefore
demonstrates that the opening main-effect procedures can detect a large
effect at this sample size; it does not directly validate the chess
opening-by-strength interaction test. We also examined where detection
fails for smaller effects. We calibrated simulations to the observed
Connect Four rates (40.6\% for the other six columns, 63.1\% for the
center) and diluted the center effect by known fractions, giving true
effects of 22.5, 16.9, 11.2, 5.6, and 2.2 points. In the same way, we
simulated the Study 1 design with a single opening whose effect at the
highest setting differs from the others. Table 4 gives the detection
rates.

\textbf{Table 4.} Detection rates of the analysis pipeline by true
effect size (percentage points), from simulation calibrated to the
observed data.

\begin{longtable}[]{@{}
  >{\raggedright\arraybackslash}p{(\columnwidth - 8\tabcolsep) * \real{0.2200}}
  >{\raggedright\arraybackslash}p{(\columnwidth - 8\tabcolsep) * \real{0.2000}}
  >{\raggedright\arraybackslash}p{(\columnwidth - 8\tabcolsep) * \real{0.2000}}
  >{\raggedright\arraybackslash}p{(\columnwidth - 8\tabcolsep) * \real{0.2000}}
  >{\raggedright\arraybackslash}p{(\columnwidth - 8\tabcolsep) * \real{0.1800}}@{}}
\toprule\noalign{}
\begin{minipage}[b]{\linewidth}\raggedright
\textbf{True effect (pp)}
\end{minipage} & \begin{minipage}[b]{\linewidth}\raggedright
\textbf{Connect Four: χ² homogeneity test}
\end{minipage} & \begin{minipage}[b]{\linewidth}\raggedright
\textbf{Connect Four: center contrast}
\end{minipage} & \begin{minipage}[b]{\linewidth}\raggedright
\textbf{Study 1: single-opening contrast (uncorrected)}
\end{minipage} & \begin{minipage}[b]{\linewidth}\raggedright
\textbf{Study 1: omnibus interaction LR test}
\end{minipage} \\
\begin{minipage}[b]{\linewidth}\raggedright
0 (no effect)
\end{minipage} & \begin{minipage}[b]{\linewidth}\raggedright
5.1\%
\end{minipage} & \begin{minipage}[b]{\linewidth}\raggedright
6.3\%
\end{minipage} & \begin{minipage}[b]{\linewidth}\raggedright
4.5\%
\end{minipage} & \begin{minipage}[b]{\linewidth}\raggedright
4.3\%
\end{minipage} \\
\begin{minipage}[b]{\linewidth}\raggedright
2.2
\end{minipage} & \begin{minipage}[b]{\linewidth}\raggedright
6.7\%
\end{minipage} & \begin{minipage}[b]{\linewidth}\raggedright
10.8\%
\end{minipage} & \begin{minipage}[b]{\linewidth}\raggedright
---
\end{minipage} & \begin{minipage}[b]{\linewidth}\raggedright
---
\end{minipage} \\
\begin{minipage}[b]{\linewidth}\raggedright
3
\end{minipage} & \begin{minipage}[b]{\linewidth}\raggedright
8.8\%
\end{minipage} & \begin{minipage}[b]{\linewidth}\raggedright
16.1\%
\end{minipage} & \begin{minipage}[b]{\linewidth}\raggedright
10.0\%
\end{minipage} & \begin{minipage}[b]{\linewidth}\raggedright
7.8\%
\end{minipage} \\
\begin{minipage}[b]{\linewidth}\raggedright
5.6
\end{minipage} & \begin{minipage}[b]{\linewidth}\raggedright
23.2\%
\end{minipage} & \begin{minipage}[b]{\linewidth}\raggedright
42.0\%
\end{minipage} & \begin{minipage}[b]{\linewidth}\raggedright
18.3\%
\end{minipage} & \begin{minipage}[b]{\linewidth}\raggedright
10.0\%
\end{minipage} \\
\begin{minipage}[b]{\linewidth}\raggedright
8
\end{minipage} & \begin{minipage}[b]{\linewidth}\raggedright
41.6\%
\end{minipage} & \begin{minipage}[b]{\linewidth}\raggedright
71.3\%
\end{minipage} & \begin{minipage}[b]{\linewidth}\raggedright
45.5\%
\end{minipage} & \begin{minipage}[b]{\linewidth}\raggedright
20.8\%
\end{minipage} \\
\begin{minipage}[b]{\linewidth}\raggedright
11.2
\end{minipage} & \begin{minipage}[b]{\linewidth}\raggedright
77.4\%
\end{minipage} & \begin{minipage}[b]{\linewidth}\raggedright
94.4\%
\end{minipage} & \begin{minipage}[b]{\linewidth}\raggedright
---
\end{minipage} & \begin{minipage}[b]{\linewidth}\raggedright
---
\end{minipage} \\
\begin{minipage}[b]{\linewidth}\raggedright
12
\end{minipage} & \begin{minipage}[b]{\linewidth}\raggedright
---
\end{minipage} & \begin{minipage}[b]{\linewidth}\raggedright
---
\end{minipage} & \begin{minipage}[b]{\linewidth}\raggedright
75.0\%
\end{minipage} & \begin{minipage}[b]{\linewidth}\raggedright
38.8\%
\end{minipage} \\
\begin{minipage}[b]{\linewidth}\raggedright
16.9
\end{minipage} & \begin{minipage}[b]{\linewidth}\raggedright
99.0\%
\end{minipage} & \begin{minipage}[b]{\linewidth}\raggedright
100.0\%
\end{minipage} & \begin{minipage}[b]{\linewidth}\raggedright
---
\end{minipage} & \begin{minipage}[b]{\linewidth}\raggedright
---
\end{minipage} \\
\begin{minipage}[b]{\linewidth}\raggedright
22.5
\end{minipage} & \begin{minipage}[b]{\linewidth}\raggedright
100.0\%
\end{minipage} & \begin{minipage}[b]{\linewidth}\raggedright
100.0\%
\end{minipage} & \begin{minipage}[b]{\linewidth}\raggedright
---
\end{minipage} & \begin{minipage}[b]{\linewidth}\raggedright
---
\end{minipage} \\
\midrule\noalign{}
\endhead
\bottomrule\noalign{}
\endlastfoot
\end{longtable}

\emph{Note.} Connect Four rows at 0, 3, and 8 points use 2,000 simulated
studies each. The other Connect Four rows use 4,000 studies each. Study
1 rows use 400 studies per effect, so Monte Carlo error is about 2 to 3
points. The Study 1 contrast uses an uncorrected 5\% test. The 5.6-point
row for Study 1 uses the 5-point simulation. With a Bonferroni
correction over ten openings, the 12-point Study 1 power falls to about
44\%. The 0-point rows show that the center contrast is slightly liberal
(6.3\%) under its normal approximation.

\includegraphics[width=6in,height=2.75in]{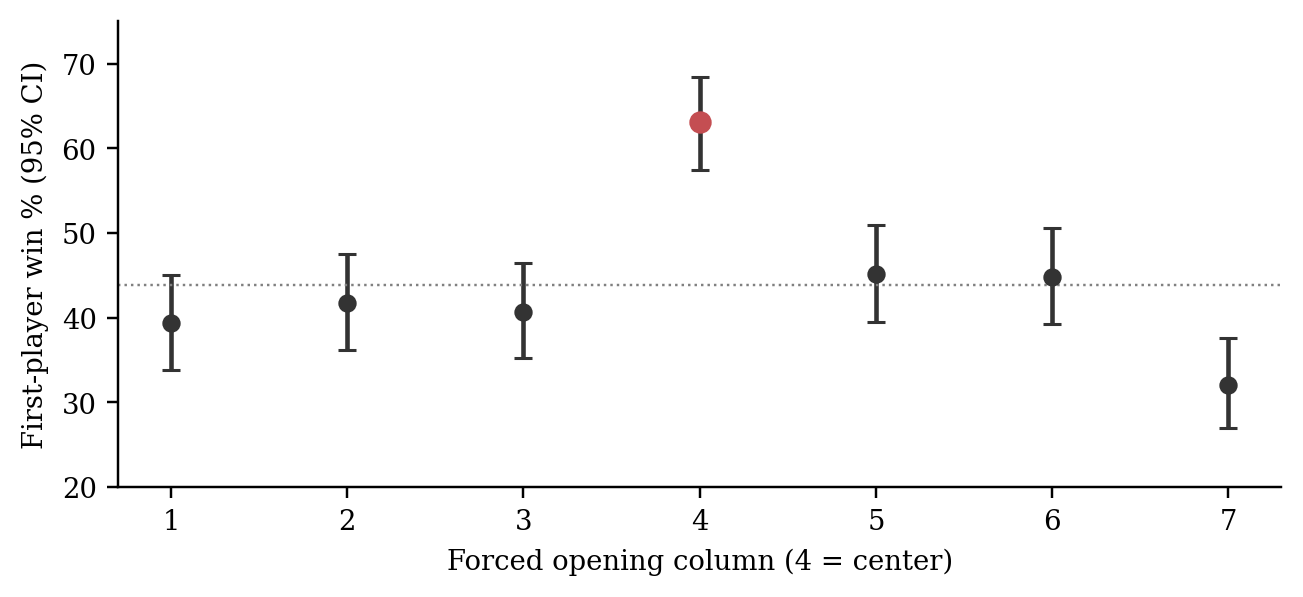}

\textbf{Figure 4.} Connect Four positive control. First-player win
percentage by forced opening column, with 95\% Wilson intervals (290
games per column, 29 runs). The center column is the first
player\textquotesingle s only winning first move under perfect play.
Full values are in Supplementary Table S20.

\subsubsection{Study 2: complete-record
replication}\label{study-2-complete-record-replication}

Study 2 did not reproduce Study 1\textquotesingle s absolute
setting-level rates (Supplementary Table S11). At 1320, draws rose from
1.1\% to 18.8\%, White wins fell from 51.0\% to 43.0\%, and Black wins
fell from 44.2\% to 35.7\%. At 2500, draws rose from 16.2\% to 24.2\%.
At 2000, the two studies were similar. The Study 2 draw rates therefore
did not rise steadily with strength. They were 18.8\%, 12.7\%, and
24.2\% at the three settings. Study 2 differed from Study 1 in both
processor and implementation (an independently written game loop,
Section 3.4). These comparisons show that absolute rates depend on the
full configuration, but they cannot isolate hardware as the cause.

The Study 2 draw pattern is unlikely to be explained by sampling noise
alone. Of the 113, 76, and 145 draws at 1320, 2000, and 2500, 46, 36,
and 72 ended in insufficient material, 22, 13, and 33 by fivefold
repetition, and 35, 23, and 30 at the 300-ply cap. The 2000 setting has
fewer draws than either neighbor. The χ² test against 1320 gives
\emph{p} = .004, and the test against 2500 gives \emph{p} \textless{}
.001. The draws are therefore mostly endgame endings, and the setting
that resolves the most games decisively is the middle one. The data show
how games end but not why. This pattern motivates a hypothesis for the
proposed confirmatory study, which will record termination reasons and
game lengths for every game.

The complete records show no repeated games, since no two shared their
first 20 plies, and draw-rate tests across the ten forced moves were not
significant after adjustment (\emph{p} from .132 to .937). Direct
estimates of White\textquotesingle s first-move advantage were 4.2
points at 1320 (95\% CI {[}−3.0, 11.4{]}), 4.3 at 2000 ({[}−3.1,
11.8{]}), and 2.5 at 2500 ({[}−4.5, 9.5{]}).

Opening effects did not differ between the two studies for White, χ²(9)
= 9.29, \emph{p} = .411, or for Black, χ²(9) = 7.10, \emph{p} = .627. No
opening-by-strength interaction was found in Study 2 alone (White
\emph{p} = .543, Black \emph{p} = .120) or in the combined data (White
\emph{p} = .407, Black \emph{p} = .680). Expected-score models gave the
same result.

\subsubsection{Study 3: validation of strength
ordering}\label{study-3-validation-of-strength-ordering}

Direct matches confirmed that each setting beat the one below it under
the study conditions (Supplementary Table S13). The 2000 setting won all
200 games against the lowest setting. The 2500 setting scored 93.8\%
against 2000, which corresponds to an estimated 470-point difference
(95\% CI {[}401, 577{]}). Against the lowest setting, however, 2500
scored 93.9\% over 400 games, an estimated 474-point difference ({[}417,
555{]}). This is far below the nominal gap of 1,150 points. The pairwise
results do not fit a single scale. Against the lowest setting, the 2000
setting scored 100\% while the 2500 setting scored only 93.9\%, so the
nominally weaker setting did better against the same opponent. The 2000
setting won all 200 of its games against the lowest setting, whereas the
2500 setting won 369 of 400, drew 13, and lost 18. All 18 losses ended
in checkmate, and they were long, averaging 155 plies. Games against the
lowest setting were also much longer at 2500 (116 plies on average) than
at 2000 (74). These long-game losses raise the possibility that game
conversion, rather than only opening play, contributed to the unexpected
pairwise scores. We cannot identify the mechanism from these data. One
candidate is the random move selection used at limited strength, which
can matter more in long games. Whatever the cause, the inconsistent
pairwise performance margins do not support interpreting the nominal
labels as calibrated distances on one Elo scale. The comparisons support
treating the settings as ordered experimental conditions, not as literal
or uniformly spaced Elo ratings.

\subsubsection{Study 4 and the separate robustness
run}\label{study-4-and-the-separate-robustness-run}

In the 2,700 White-opening games played after the 1.d3 hypothesis was
fixed, 1.d3 won 37.4\% of its 270 games, compared with 40.0\% for the
other moves (Supplementary Table S14). The prespecified one-sided
Mantel--Haenszel test did not reach the threshold: common odds ratio
0.90, 95\% CI {[}0.69, 1.16{]}, \emph{p} = .208. Expected scores were
46.5\% and 49.7\% (\emph{p} = .134). The Study 1 deficit was therefore
not confirmed in new games, although the new estimate pointed in the
same direction. The interval includes both no effect and the Study 1
estimate. The follow-up therefore neither confirms nor rules out a
deficit as large as the one in Study 1, and it does not show that 1.d3
is equivalent to the other moves. Pooling these data with the separate
robustness run gave a similar odds ratio of 0.91 (95\% CI {[}0.75,
1.09{]}, one-sided \emph{p} = .146), but that combined analysis was not
prespecified.

The separate robustness run gave setting-level rates within 4.2 points
of the laptop run, found no interaction (White and Black \emph{p} =
.521), and did not confirm the 1.d3 deficit (odds ratio 0.91, one-sided
\emph{p} = .249). Because both follow-up runs used the same program, it
is a robustness check rather than an independent software replication.

\subsection{4.5 Fixed-compute
extensions}\label{fixed-compute-extensions}

\subsubsection{Study 5: independent evaluation of selected
replies}\label{study-5-independent-evaluation-of-selected-replies}

The Study 5 survey produced 6,955 completed games, and the independent
evaluation produced 5,216 of 5,220 scheduled games, with censored games
excluded. The learned policy chose different replies by opponent
strength after 1.e4 and 1.d4, and chose 1...c5 at both settings after
1.c4 (Supplementary Table S15). Several leading survey estimates were
close, so the selected replies were themselves uncertain.

Replies selected from the survey showed no statistically reliable
advantage over letting Stockfish choose freely (Figure 5 and
Supplementary Table S16). Against White at 1800, the learned
policy\textquotesingle s Black expected score was 3.32 points lower than
the free engine\textquotesingle s (95\% CI {[}−7.38, +0.74{]}). Against
White at 2000, it was 3.93 points lower ({[}−8.69, +0.84{]}). Both
Holm-adjusted \emph{p} values were .213. The fixed book did not differ
detectably from the free engine, and all six opening-specific estimates
were negative but included zero (Supplementary Table S17). The study
therefore neither shows that selection helps nor shows that it harms.

\includegraphics[width=6in,height=2.53125in]{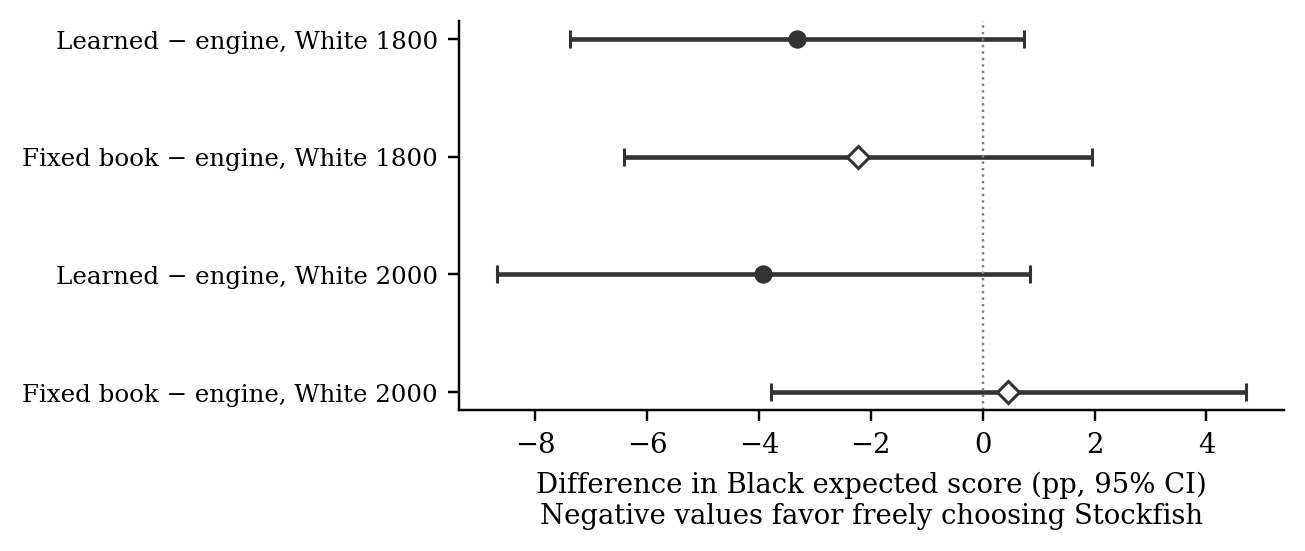}

\textbf{Figure 5.} Study 5 differences in Black expected score relative
to letting Stockfish choose freely, with block-level 95\% confidence
intervals. Negative values favor the free engine.

\subsubsection{Study 6: forced multi-move
variations}\label{study-6-forced-multi-move-variations-1}

Study 6 recorded 10,434 completed games of 10,440 scheduled (3,679 White
wins, 4,881 Black wins, and 1,874 draws, with six censored).
Black\textquotesingle s mean expected score was 64.55\% against White at
1800 and 46.97\% against White at 2000 (Figure 6 and Supplementary Table
S18). This 17.6-point shift with White\textquotesingle s setting
dominates the figure. Seven repeated move sequences in two Caro-Kann
Gurgenidze cells changed that variation\textquotesingle s score by less
than one point.

\includegraphics[width=6in,height=5.70833in]{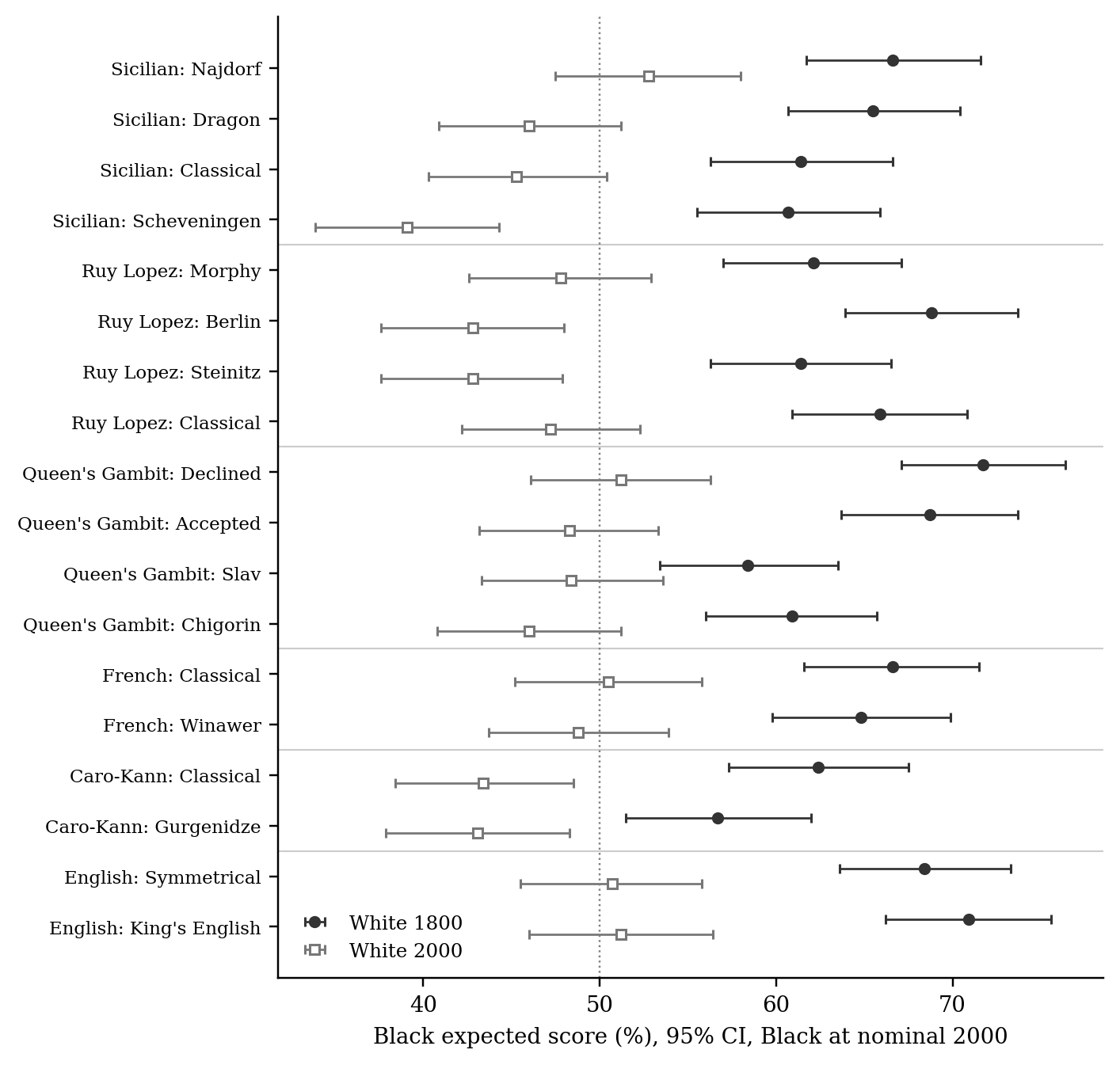}

\textbf{Figure 6.} Study 6 Black expected score for 18 forced variations
against White at nominal 1800 and 2000, with descriptive 95\% confidence
intervals (290 scheduled games per point; six games censored overall).
Horizontal rules separate opening families, and the dotted line marks
50\%.

Two of the 42 exploratory within-family comparisons survived Holm
adjustment (Supplementary Table S19). Against White at 2000, the Najdorf
Sicilian scored 13.68 points above the Scheveningen (adjusted \emph{p} =
.0013, multiplicity-adjusted 99.88\% CI {[}+3.73, +23.62{]}). Against
White at 1800, the Queen\textquotesingle s Gambit Declined scored 13.28
points above the Slav (adjusted \emph{p} = .0053, 99.88\% CI {[}+2.49,
+24.07{]}). These estimates are larger than the estimated differences
among first moves in Studies 1 and 2. They are, however, the largest of
42 exploratory comparisons and are therefore likely to overstate the
true differences. Forced multi-move lines are also a different kind of
treatment from single first moves.

No variation-by-strength interaction survived adjustment across the 21
tests, and the smallest adjusted \emph{p}-value was .433. Study 6
therefore finds two exploratory differences among complete variations
that survive adjustment, but no statistically reliable evidence that
their relative performance changes with the opponent\textquotesingle s
strength.

\section{5. Discussion}\label{discussion}

\subsection{5.1 Principal finding}\label{principal-finding}

The leading opening changed across strength settings, but there was no
statistically reliable evidence that the best opening changed with
strength. Neither interaction test was significant for either color, no
opening-specific change survived correction, and the follow-up studies
agree. The Connect Four control and the sensitivity analysis show that
the pipeline finds large effects, but that effects of a few points would
often be missed (Table 4). The changes in leaders therefore do not
establish strength dependence, and modest interactions remain possible.

\subsection{5.2 Why rankings and win rates
mislead}\label{why-rankings-and-win-rates-mislead}

The observed leader is sensitive to sampling error. At a between-opening
standard deviation of three points, the observed leader after 290 games
per opening was truly best in only about 44\% of simulated studies
(Figure 3). A ranking is therefore an estimate with uncertainty, not a
directly observed property of an opening. Win rates also move with draw
rates. In Study 1, draws rose sharply at the higher settings, so wins
for both colors fell together while expected score changed little.
Absolute rates also changed on different hardware and with a different
game loop (Study 2), so they should be read as properties of the full
configuration.

\subsection{5.3 Selection, complexity, and the 1.d3
result}\label{selection-complexity-and-the-1.d3-result}

Replies selected because they scored best in the survey did no better
than letting Stockfish choose freely in new games. This is the pattern
expected from choosing the maximum of several noisy estimates. The
survey did not include the free-engine comparison, however, so other
explanations remain possible. Selection and evaluation data should
always be kept separate when an experiment is used to recommend an
opening.

The complexity hypothesis is not tested here, because the completed
studies found no reliable pattern of changing relative performance for
complexity to explain. Classifying openings as complex after seeing
their rankings would be circular, so a valid test needs a measure
defined in advance (Section 7). The 1.d3 result shows the same need. Its
Study 1 deficit of about seven points shrank to about two to three
points in new games, below the prespecified threshold.

\subsection{5.4 What this means for choosing an
opening}\label{what-this-means-for-choosing-an-opening}

The study does not identify an opening that suits a particular level,
and it gives no basis for recommending a first move or reply to players
of any rating. The leader at each setting changed within the range that
sampling noise produces, and the best survey replies did no better than
free engine choice. The settings are also not human ratings (Section
6.4). Even a reliable engine-level interaction would describe Stockfish,
not club or tournament players.

One possible extension is that humans could adopt the
engine\textquotesingle s tendencies, meaning the openings the engine
scores well with at a given setting. This is untested, and there are
reasons for caution. Human players study some openings intensively and
neglect others, so an opening the engine favors may be unfamiliar to
people, or already heavily analyzed and well defended. Engine tendencies
also did not transfer even within the engine, as Section 5.3 shows.
Testing the idea would require human games with controls for preparation
and selection, such as randomized assignment of openings in a controlled
setting or a carefully matched comparison of human results. Until such a
test supports them, claims of a level-specific best opening should be
treated with caution.

\section{6. Limitations}\label{limitations}

\subsection{6.1 Data provenance}\label{data-provenance}

Study 1 was reconstructed from screenshots of summary displays rather
than from game records. The per-cell totals matched independently
transcribed summaries exactly, which makes a large systematic error
unlikely, but errors that cancel out among openings within a run cannot
be excluded. Study 1 also lacks move sequences, per-opening draw counts,
and termination reasons. Study 2 supplies complete records under the
same software settings, but it is much smaller and ran on different
hardware with an independently written game loop.

\subsection{6.2 Strength conditions}\label{strength-conditions}

The nominal labels are ordered experimental settings, not calibrated Elo
ratings. Stockfish 15.1 calibrates UCI\_Elo under much longer time
controls than the 10 milliseconds per move used here, so actual strength
was probably below the nominal values, and the gaps need not match the
labels. The platform also departed from standard play: the engine
received the position without the move history, and fivefold rather than
claimable threefold repetition applied. These choices may have changed
game length and draw frequency.

\subsection{6.3 Statistical and confirmatory
limits}\label{statistical-and-confirmatory-limits}

The design has limited sensitivity to small interactions (Table 4).
Multiplicity adjustments reduce false positives within families of tests
but do not turn exploratory selection into confirmation. The Study 4
hypothesis was fixed in an earlier version of this manuscript but was
not publicly registered. The reply sets and variations in Studies 5 and
6 were chosen by the authors, so those contrasts remain exploratory.

\subsection{6.4 Scope}\label{scope}

The findings apply to the tested Stockfish releases, strength
mechanisms, search budgets, hardware, and forced treatments. Nominal
settings are not measures of human skill, so the results should not be
applied directly to human players. Studies 1--5 forced only a first move
or reply, and Study 6 covered 18 sequences. Studies 5 and 6 used a newer
engine release, a node budget, different draw rules, and an asymmetric
strength design, so they extend rather than replicate the original
evidence.

\section{7. Proposed Confirmatory
Study}\label{proposed-confirmatory-study}

\subsection{7.1 Rationale}\label{rationale}

The completed studies leave small strength-dependent effects unresolved.
The proposed confirmatory study tests a prespecified opening-by-strength
interaction, evaluates a complexity measure defined without reference to
outcomes, and reassesses the 1.d3 result under a reproducible
configuration. It will also include a real small-effect check, in which
known handicaps are implanted in actual games, to confirm that the
analysis detects effects of the size predicted by the sensitivity
simulations. No data from this study have been collected or analyzed.
The protocol will be registered on OSF Registries before any
confirmatory games are generated.

\subsection{7.2 Planned hypotheses}\label{planned-hypotheses}

\textbf{Table 5.} Planned hypotheses for the proposed confirmatory
study.

\begin{longtable}[]{@{}
  >{\raggedright\arraybackslash}p{(\columnwidth - 6\tabcolsep) * \real{0.1599}}
  >{\raggedright\arraybackslash}p{(\columnwidth - 6\tabcolsep) * \real{0.2799}}
  >{\raggedright\arraybackslash}p{(\columnwidth - 6\tabcolsep) * \real{0.2599}}
  >{\raggedright\arraybackslash}p{(\columnwidth - 6\tabcolsep) * \real{0.3002}}@{}}
\toprule\noalign{}
\begin{minipage}[b]{\linewidth}\raggedright
\textbf{Hypothesis}
\end{minipage} & \begin{minipage}[b]{\linewidth}\raggedright
\textbf{Estimand}
\end{minipage} & \begin{minipage}[b]{\linewidth}\raggedright
\textbf{Planned analysis}
\end{minipage} & \begin{minipage}[b]{\linewidth}\raggedright
\textbf{Status before registration}
\end{minipage} \\
\begin{minipage}[b]{\linewidth}\raggedright
H1: Strength dependence
\end{minipage} & \begin{minipage}[b]{\linewidth}\raggedright
Change in relative opening performance between the 1320 and 2500
settings, by color
\end{minipage} & \begin{minipage}[b]{\linewidth}\raggedright
Opening-by-strength interaction for designated-color expected score
\end{minipage} & \begin{minipage}[b]{\linewidth}\raggedright
Primary. The model, test, and alpha allocation across colors are still
to be fixed.
\end{minipage} \\
\begin{minipage}[b]{\linewidth}\raggedright
H2: Complexity
\end{minipage} & \begin{minipage}[b]{\linewidth}\raggedright
Association between each opening\textquotesingle s interaction contrast
and its complexity index
\end{minipage} & \begin{minipage}[b]{\linewidth}\raggedright
One-sided Spearman permutation test, permuted within color
\end{minipage} & \begin{minipage}[b]{\linewidth}\raggedright
Primary. The index and its power are still to be specified.
\end{minipage} \\
\begin{minipage}[b]{\linewidth}\raggedright
H3: 1.d3 replication
\end{minipage} & \begin{minipage}[b]{\linewidth}\raggedright
Common deficit of 1.d3 relative to the other nine White moves
\end{minipage} & \begin{minipage}[b]{\linewidth}\raggedright
One-sided stratified test
\end{minipage} & \begin{minipage}[b]{\linewidth}\raggedright
The primary outcome, and whether the claim is common or
setting-specific, are still to be fixed.
\end{minipage} \\
\begin{minipage}[b]{\linewidth}\raggedright
H4: Draw-rate ordering
\end{minipage} & \begin{minipage}[b]{\linewidth}\raggedright
Draw-rate differences for 2000 versus 1320 and 2500 versus 2000
\end{minipage} & \begin{minipage}[b]{\linewidth}\raggedright
Two one-sided comparisons
\end{minipage} & \begin{minipage}[b]{\linewidth}\raggedright
Confirmatory status must be reconciled with the smaller middle-setting
sample.
\end{minipage} \\
\midrule\noalign{}
\endhead
\bottomrule\noalign{}
\endlastfoot
\end{longtable}

\emph{Note.} Entries marked as still to be resolved are not confirmatory
specifications until they are finalized in the registration.

\subsection{7.3 Design and sample size}\label{design-and-sample-size}

The design keeps the ten White first moves and ten Black replies. It
uses 2,100 games per opening-by-setting cell at the lowest and highest
settings, and 700 at the middle setting. For an 8-point interaction
contrast, about 1,215 games per cell give 80\% power at α = .05, and
about 2,060 at α = .005 (Figure 7). The complete design requires 98,000
games. Games will be randomized, each will have a logged seed, and the
archive will keep complete PGN records, termination reasons, and ply
counts.

\includegraphics[width=6in,height=2.94792in]{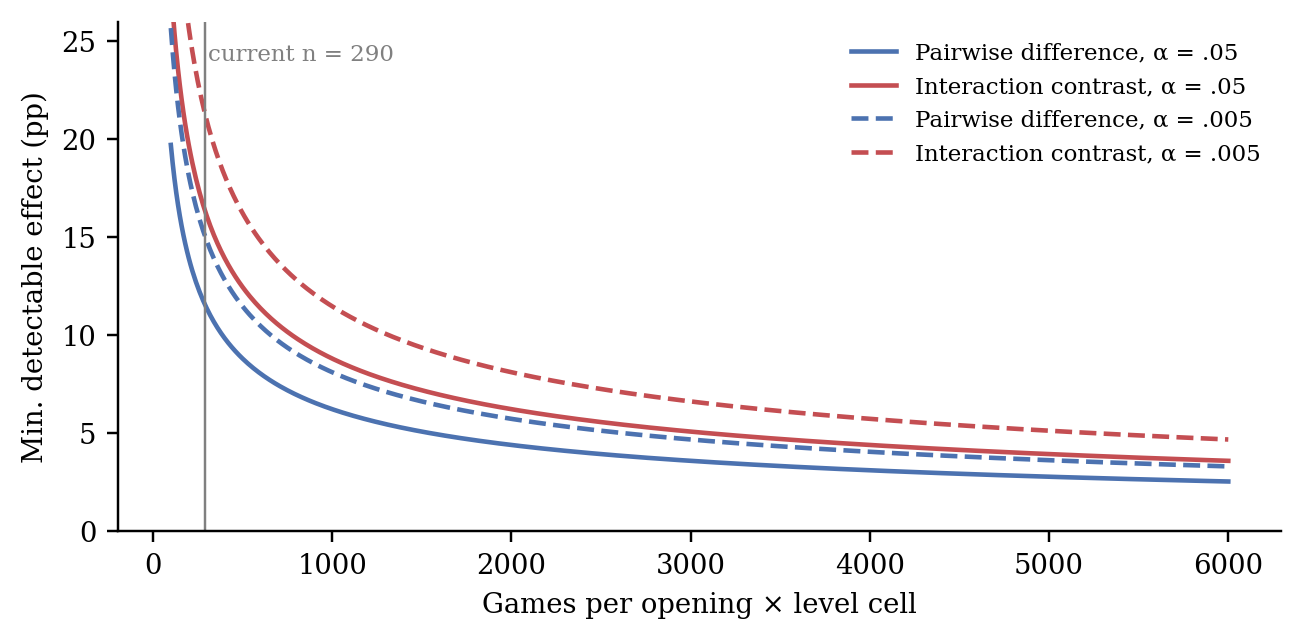}

\textbf{Figure 7.} Minimum detectable effect at 80\% power as a function
of games per opening-by-setting cell. Solid lines use α = .05 and dashed
lines use α = .005. The vertical line marks the current 290-game design.

\subsection{7.4 Complexity index and
analysis}\label{complexity-index-and-analysis}

For each forced move, 500 positions will be sampled from full-strength
continuations and analyzed at a fixed node budget. The index combines
breadth (the number of moves within 50 centipawns of the best move),
narrowness (the proportion of positions with only one such move), and
cost (the second-best move\textquotesingle s evaluation loss, capped at
300 centipawns). It is standardized as (\emph{z}\textsubscript{narrow} +
\emph{z}\textsubscript{cost} − \emph{z}\textsubscript{breadth}) / 3. The
primary model, the multiplicity correction, and the exclusion rules will
be fixed in the registration. A run will be excluded only for a
documented technical failure identified without looking at its outcomes.

\section{8. Conclusion}\label{conclusion}

Across 17,400 strength-limited Stockfish games, a Connect Four positive
control, replication and validation studies, and two fixed-compute
extensions, the observed opening leaders changed more readily than the
estimated differences among openings. No completed analysis found a
statistically reliable opening-by-strength interaction. That does not
show that all interactions are absent, and the sensitivity analysis
shows that the design would often miss effects of a few points. The
exploratory 1.d3 deficit was not confirmed in new games, and replies
selected from one set of engine games did no better than free engine
choice. A leaderboard built from a few hundred games should be treated
as a noisy estimate, not a stable hierarchy. Reporting uncertainty,
testing interactions directly, keeping complete records, documenting the
full configuration, and separating selection from evaluation data would
make future engine-opening comparisons more credible.

\section{Declarations}\label{declarations}

\subsection{Data and code
availability}\label{data-and-code-availability}

The archived record at Zenodo (Flick et al., 2026,
https://doi.org/10.5281/zenodo.23357406) contains the transcribed Study
1 data (cells.csv and run\_summaries.csv), the Connect Four data, the
original screenshots, the replication program, complete game records for
Studies 2--4 and the robustness run in JSON Lines format with move
lists, game-level CSV and PGN records for Studies 5 and 6, protocols,
schedules, analysis outputs, and the analysis and figure scripts. The
source code is also on GitHub at
https://github.com/FLICK11-hash/gaming-platform. Supplementary Table S20
gives the Connect Four column counts.

\subsection{Author contributions}\label{author-contributions}

Conrad Flick: Conceptualization, Methodology, Software, Investigation,
Data curation, Formal analysis, Visualization, Writing -- original
draft, and Writing -- review \& editing. Jaysen Jensen: Writing --
review \& editing, including editorial restructuring and revision for
clarity and accessibility. Jingyu Huang: Methodology, Software,
Investigation, Formal analysis, Data curation, and Writing -- original
draft for Studies 5 and 6. Mario Camara Paslar: Writing -- review \&
editing, including revision of the manuscript and recommendations about
its presentation and interpretation.

\subsection{Funding}\label{funding}

This research received no specific grant from any funding agency in the
public, commercial, or not-for-profit sectors.

\subsection{Conflicts of interest}\label{conflicts-of-interest}

The authors declare no potential conflicts of interest with respect to
the research, authorship, or publication of this article.

\subsection{Acknowledgments}\label{acknowledgments}

The authors thank Carey Main of the University of Chicago, who mentored
Conrad Flick through a University of Chicago summer mentorship, for her
guidance, for suggestions that shaped parts of the project, and for
contributing portions of the simulation platform code.

\subsection{AI-assisted tools
disclosure}\label{ai-assisted-tools-disclosure}

Claude (Anthropic) was used to assist with grammar and wording, figure
preparation, and code generation. Its code contributions included
portions of the simulation platform, as recorded in the
repository\textquotesingle s commit history, and the replication program
and statistical analysis scripts. Claude also assisted with drafting and
revising portions of the manuscript. The independent robustness run was
executed by Claude on a cloud processor and is reported separately from
the primary analyses. Codex (OpenAI) was used during an earlier revision
to assist with editorial restructuring, language revision, HTML
formatting, and the placement of tables and figures. In the most recent
revisions, Claude shortened the manuscript, drafted the human-copying
paragraph in Section 5.4, wrote the sensitivity simulations and the
explanations of the Study 2 and Study 3 patterns, restored the Connect
Four control, rebuilt the tables, regenerated the supplementary exhibits
from the archived scripts, and prepared the LaTeX source. The authors
conducted all primary experiments and take full responsibility for the
content of the article.

\section{References}\label{references}

Agresti, A. (2013). \emph{Categorical data analysis} (3rd ed.). Wiley.

Allis, L. V. (1988). \emph{A knowledge-based approach of Connect-Four:
The game is solved: White wins} {[}Master\textquotesingle s thesis,
Vrije Universiteit Amsterdam{]}.

Anderson, A., Kleinberg, J., \& Mullainathan, S. (2017). Assessing human
error against a benchmark of perfection. \emph{ACM Transactions on
Knowledge Discovery from Data, 11}(4), Article 45.

Benjamini, Y., \& Hochberg, Y. (1995). Controlling the false discovery
rate: A practical and powerful approach to multiple testing.
\emph{Journal of the Royal Statistical Society: Series B
(Methodological), 57}(1), 289--300.

Chase, W. G., \& Simon, H. A. (1973). Perception in chess.
\emph{Cognitive Psychology, 4}(1), 55--81.

de Groot, A. D. (1965). \emph{Thought and choice in chess}. Mouton.

Efron, B., \& Tibshirani, R. J. (1993). \emph{An introduction to the
bootstrap}. Chapman \& Hall.

Ferreira, D. R. (2013). The impact of the search depth on chess playing
strength. \emph{ICGA Journal, 36}(2), 67--80.

Fiekas, N. (n.d.). \emph{python-chess} (Version 1.11.2) {[}Computer
software{]}. https://github.com/niklasf/python-chess

Flick, C., Camara Paslar, M., \& Huang, J. (2026).
\emph{gaming-platform: Simulation platform, data, and analysis code for
opening rankings in engine self-play} {[}Computer software and data
set{]}. Zenodo. https://doi.org/10.5281/zenodo.23357406

Gelman, A., \& Loken, E. (2014). The statistical crisis in science.
\emph{American Scientist, 102}(6), 460--465.

Gobet, F., \& Simon, H. A. (1996). Templates in chess memory: A
mechanism for recalling several boards. \emph{Cognitive Psychology,
31}(1), 1--40.

Guid, M., \& Bratko, I. (2006). Computer analysis of world chess
champions. \emph{ICGA Journal, 29}(2), 65--73.

Holm, S. (1979). A simple sequentially rejective multiple test
procedure. \emph{Scandinavian Journal of Statistics, 6}(2), 65--70.

Kerr, N. L. (1998). HARKing: Hypothesizing after the results are known.
\emph{Personality and Social Psychology Review, 2}(3), 196--217.

Mantel, N., \& Haenszel, W. (1959). Statistical aspects of the analysis
of data from retrospective studies of disease. \emph{Journal of the
National Cancer Institute, 22}(4), 719--748.

McIlroy-Young, R., Sen, S., Kleinberg, J., \& Anderson, A. (2020).
Aligning superhuman AI with human behavior: Chess as a model system. In
\emph{Proceedings of the 26th ACM SIGKDD International Conference on
Knowledge Discovery \& Data Mining} (pp. 1677--1687). ACM.

Nosek, B. A., Ebersole, C. R., DeHaven, A. C., \& Mellor, D. T. (2018).
The preregistration revolution. \emph{Proceedings of the National
Academy of Sciences, 115}(11), 2600--2606.

notnil. (n.d.). \emph{chess} (Version 1.10.0) {[}Computer software{]}.
https://github.com/notnil/chess

Regan, K. W., \& Haworth, G. M. (2011). Intrinsic chess ratings. In
\emph{Proceedings of the Twenty-Fifth AAAI Conference on Artificial
Intelligence} (pp. 834--839). AAAI Press.

Seabold, S., \& Perktold, J. (2010). Statsmodels: Econometric and
statistical modeling with Python. In \emph{Proceedings of the 9th Python
in Science Conference} (pp. 92--96).

Silver, D., Hubert, T., Schrittwieser, J., Antonoglou, I., Lai, M.,
Guez, A., Lanctot, M., Sifre, L., Kumaran, D., Graepel, T., Lillicrap,
T., Simonyan, K., \& Hassabis, D. (2018). A general reinforcement
learning algorithm that masters chess, shogi, and Go through self-play.
\emph{Science, 362}(6419), 1140--1144.

Stockfish developers. (2022). \emph{Stockfish} (Version 15.1)
{[}Computer software{]}. https://stockfishchess.org

Stockfish developers. (2026a). \emph{UCI protocol and Stockfish
commands} {[}Software documentation{]}.
https://official-stockfish.github.io/docs/stockfish-wiki/UCI-Protocol-and-Stockfish-Commands.html

Stockfish developers. (2026b). \emph{Stockfish} (Version 19) {[}Computer
software{]}. https://stockfishchess.org

Tomašev, N., Paquet, U., Hassabis, D., \& Kramnik, V. (2020).
\emph{Assessing game balance with AlphaZero: Exploring alternative rule
sets in chess} (arXiv:2009.04374). arXiv.

Virtanen, P., Gommers, R., Oliphant, T. E., Haberland, M., Reddy, T.,
Cournapeau, D., et al. (2020). SciPy 1.0: Fundamental algorithms for
scientific computing in Python. \emph{Nature Methods, 17}, 261--272.

Wilson, E. B. (1927). Probable inference, the law of succession, and
statistical inference. \emph{Journal of the American Statistical
Association, 22}(158), 209--212.

\section{Appendix A. Derivation of the First-Move Advantage
Bounds}\label{appendix-a.-derivation-of-the-first-move-advantage-bounds}

Consider one strength setting, with 100 games in the White-opening set
(set 1) and 100 games in the Black-reply set (set 2). Let \emph{W} be
White\textquotesingle s win percentage in set 1, \emph{B}
Black\textquotesingle s win percentage in set 2, and
\emph{d}\textsubscript{1} and \emph{d}\textsubscript{2} the draw
percentages in the two sets. Every game ends in a White win, a Black
win, or a draw. Black therefore wins 100 − \emph{W} −
\emph{d}\textsubscript{1} percent of set 1, and White wins 100 −
\emph{B} − \emph{d}\textsubscript{2} percent of set 2.

Across the pooled 200 games, White\textquotesingle s win percentage is
{[}\emph{W} + (100 − \emph{B} − \emph{d}\textsubscript{2}){]} / 2 and
Black\textquotesingle s is {[}\emph{B} + (100 − \emph{W} −
\emph{d}\textsubscript{1}){]} / 2. Their difference is

\emph{Δ = W − B + (d₁ − d₂)/2.}

If the displayed draw rate \emph{D} is the draw percentage over all 200
games, then \emph{D} = (\emph{d}\textsubscript{1} +
\emph{d}\textsubscript{2}) / 2, with \emph{d}\textsubscript{1} ≥ 0 and
\emph{d}\textsubscript{2} ≥ 0. The term (\emph{d}\textsubscript{1} −
\emph{d}\textsubscript{2}) / 2 is largest when \emph{d}\textsubscript{2}
= 0 and \emph{d}\textsubscript{1} = 2\emph{D}, and smallest in the
reverse case. Therefore

\emph{W − B − D ≤ Δ ≤ W − B + D.}

When \emph{d}\textsubscript{1} = \emph{d}\textsubscript{2}, Δ = \emph{W}
− \emph{B}. White\textquotesingle s expected score in the pooled games
is White\textquotesingle s win percentage plus \emph{D} / 2. This
simplifies to (\emph{W} + 100 − \emph{B}) / 2 +
(\emph{d}\textsubscript{1} − \emph{d}\textsubscript{2}) / 4, and it
therefore lies within (\emph{W} + 100 − \emph{B}) / 2 ± \emph{D} / 2.

Averaging these bounds across runs gives the values in Supplementary
Table S9. Because the bounds are linear in \emph{W}, \emph{B}, and
\emph{D}, the bound for the mean equals the mean of the run-specific
bounds. Tighter bounds could be obtained by capping
\emph{d}\textsubscript{1} and \emph{d}\textsubscript{2} based on the
observed spread of run-level draw rates, but the analyses do not rely on
such caps.

\section{Supplementary Material}\label{supplementary-material}

The supporting research package is available through the project
repository and the permanent Zenodo archive, which is the version of
record for the deposited materials. It contains the Study 1
transcriptions (cells.csv and run\_summaries.csv) and original simulator
screenshots, the replication program and engine-build instructions,
complete Study 2--4 and robustness-run records in JSON Lines format with
move lists, Study 5--6 game-level CSV and PGN records with protocols,
schedules, and analysis outputs, and the analysis and figure scripts.

Supplementary Tables S1--S19 follow the order of the paper, from
configuration and design through the six completed studies. Table S20
gives the Connect Four positive-control data. Supplementary Figures
S1--S7 follow in the same order. Exhibits that already appear in the
main text or appendices are not repeated here. Values in Tables S8--S14
were produced by the archived analysis scripts, values in Tables
S15--S18 come from the archived Study 5 and 6 analysis outputs, and
Table S20 comes from the archived Connect Four analysis.

\subsection{Supplementary Tables}\label{supplementary-tables}

\textbf{Table S1.} Engine and simulator configuration used for all Study
1 runs.



\emph{Note.} Values are taken from the platform source code
(internal/engine/stockfish, internal/domain/research,
internal/research/simulate, and the Dockerfile). The chess library does
not apply threefold repetition or the 50-move rule automatically, so
games can continue past them until one of the listed conditions is met.

\textbf{Table S2.} Forced first moves and the labels used by the
simulator.

%

\emph{Note.} In the Black-reply set, every game began with 1.e4 followed
by the forced reply. Labels appear in parentheses in the text for
continuity with the simulator output.

\textbf{Table S3.} Outcomes by nominal strength setting, estimated from
29 run-level summaries.

%

\emph{Note.} Bracketed values are 95\% t intervals across runs. The 2000
draw-rate estimate uses 28 runs because the Run 12 summary was
unavailable. \textsuperscript{a}Implied values that assume equal draw
rates in the two sets at a setting. Appendix A gives bounds that do not
require this assumption.

\textbf{Table S4.} Heterogeneity among the ten openings of each color
within each strength setting.

\begin{longtable}[]{@{}
  >{\raggedright\arraybackslash}p{(\columnwidth - 20\tabcolsep) * \real{0.0878}}
  >{\raggedright\arraybackslash}p{(\columnwidth - 20\tabcolsep) * \real{0.0878}}
  >{\raggedright\arraybackslash}p{(\columnwidth - 20\tabcolsep) * \real{0.0878}}
  >{\raggedright\arraybackslash}p{(\columnwidth - 20\tabcolsep) * \real{0.0878}}
  >{\raggedright\arraybackslash}p{(\columnwidth - 20\tabcolsep) * \real{0.0769}}
  >{\raggedright\arraybackslash}p{(\columnwidth - 20\tabcolsep) * \real{0.0988}}
  >{\raggedright\arraybackslash}p{(\columnwidth - 20\tabcolsep) * \real{0.0988}}
  >{\raggedright\arraybackslash}p{(\columnwidth - 20\tabcolsep) * \real{0.0988}}
  >{\raggedright\arraybackslash}p{(\columnwidth - 20\tabcolsep) * \real{0.0878}}
  >{\raggedright\arraybackslash}p{(\columnwidth - 20\tabcolsep) * \real{0.0878}}
  >{\raggedright\arraybackslash}p{(\columnwidth - 20\tabcolsep) * \real{0.0997}}@{}}
\toprule\noalign{}
\begin{minipage}[b]{\linewidth}\raggedright
\textbf{Color}
\end{minipage} & \begin{minipage}[b]{\linewidth}\raggedright
\textbf{Setting}
\end{minipage} & \begin{minipage}[b]{\linewidth}\raggedright
\textbf{Pooled win \%}
\end{minipage} & \begin{minipage}[b]{\linewidth}\raggedright
\textbf{χ²(9)}
\end{minipage} & \begin{minipage}[b]{\linewidth}\raggedright
\textbf{p}
\end{minipage} & \begin{minipage}[b]{\linewidth}\raggedright
\textbf{Observed range (pp)}
\end{minipage} & \begin{minipage}[b]{\linewidth}\raggedright
\textbf{Null median range}
\end{minipage} & \begin{minipage}[b]{\linewidth}\raggedright
\textbf{Null 95th pct.}
\end{minipage} & \begin{minipage}[b]{\linewidth}\raggedright
\textbf{p (range)}
\end{minipage} & \begin{minipage}[b]{\linewidth}\raggedright
\textbf{τ (pp)}
\end{minipage} & \begin{minipage}[b]{\linewidth}\raggedright
\textbf{Reliability}
\end{minipage} \\
\end{longtable}

\emph{Note.} Null distributions of the range come from 100,000
simulations with equal win probabilities. τ is the method-of-moments
estimate of the between-opening standard deviation, and reliability is
τ²/(τ² + σ²). Holm-adjusted p-values for the six χ² tests are all ≥ .17.

\textbf{Table S5.} White first moves: win percentages, absolute change
from 1320 to 2500, and difference-in-differences (DiD) relative to the
other nine moves.

\begin{longtable}[]{@{}
  >{\raggedright\arraybackslash}p{(\columnwidth - 16\tabcolsep) * \real{0.2210}}
  >{\raggedright\arraybackslash}p{(\columnwidth - 16\tabcolsep) * \real{0.0736}}
  >{\raggedright\arraybackslash}p{(\columnwidth - 16\tabcolsep) * \real{0.0736}}
  >{\raggedright\arraybackslash}p{(\columnwidth - 16\tabcolsep) * \real{0.0736}}
  >{\raggedright\arraybackslash}p{(\columnwidth - 16\tabcolsep) * \real{0.1684}}
  >{\raggedright\arraybackslash}p{(\columnwidth - 16\tabcolsep) * \real{0.0736}}
  >{\raggedright\arraybackslash}p{(\columnwidth - 16\tabcolsep) * \real{0.1684}}
  >{\raggedright\arraybackslash}p{(\columnwidth - 16\tabcolsep) * \real{0.0736}}
  >{\raggedright\arraybackslash}p{(\columnwidth - 16\tabcolsep) * \real{0.0741}}@{}}
\toprule\noalign{}
\begin{minipage}[b]{\linewidth}\raggedright
\textbf{Move}
\end{minipage} & \begin{minipage}[b]{\linewidth}\raggedright
\textbf{1320}
\end{minipage} & \begin{minipage}[b]{\linewidth}\raggedright
\textbf{2000}
\end{minipage} & \begin{minipage}[b]{\linewidth}\raggedright
\textbf{2500}
\end{minipage} & \begin{minipage}[b]{\linewidth}\raggedright
\textbf{Change {[}95\% CI{]}}
\end{minipage} & \begin{minipage}[b]{\linewidth}\raggedright
\textbf{Holm p}
\end{minipage} & \begin{minipage}[b]{\linewidth}\raggedright
\textbf{DiD {[}95\% CI{]}}
\end{minipage} & \begin{minipage}[b]{\linewidth}\raggedright
\textbf{p}
\end{minipage} & \begin{minipage}[b]{\linewidth}\raggedright
\textbf{Holm p}
\end{minipage} \\
\end{longtable}

\emph{Note.} Values are percentages or percentage points. Rows are
ordered by DiD. Confidence intervals are unadjusted.

\textbf{Table S6.} Black first replies: win percentages, absolute change
from 1320 to 2500, and difference-in-differences (DiD) relative to the
other nine replies.

\begin{longtable}[]{@{}
  >{\raggedright\arraybackslash}p{(\columnwidth - 16\tabcolsep) * \real{0.2210}}
  >{\raggedright\arraybackslash}p{(\columnwidth - 16\tabcolsep) * \real{0.0736}}
  >{\raggedright\arraybackslash}p{(\columnwidth - 16\tabcolsep) * \real{0.0736}}
  >{\raggedright\arraybackslash}p{(\columnwidth - 16\tabcolsep) * \real{0.0736}}
  >{\raggedright\arraybackslash}p{(\columnwidth - 16\tabcolsep) * \real{0.1684}}
  >{\raggedright\arraybackslash}p{(\columnwidth - 16\tabcolsep) * \real{0.0736}}
  >{\raggedright\arraybackslash}p{(\columnwidth - 16\tabcolsep) * \real{0.1684}}
  >{\raggedright\arraybackslash}p{(\columnwidth - 16\tabcolsep) * \real{0.0736}}
  >{\raggedright\arraybackslash}p{(\columnwidth - 16\tabcolsep) * \real{0.0741}}@{}}
\toprule\noalign{}
\begin{minipage}[b]{\linewidth}\raggedright
\textbf{Reply}
\end{minipage} & \begin{minipage}[b]{\linewidth}\raggedright
\textbf{1320}
\end{minipage} & \begin{minipage}[b]{\linewidth}\raggedright
\textbf{2000}
\end{minipage} & \begin{minipage}[b]{\linewidth}\raggedright
\textbf{2500}
\end{minipage} & \begin{minipage}[b]{\linewidth}\raggedright
\textbf{Change {[}95\% CI{]}}
\end{minipage} & \begin{minipage}[b]{\linewidth}\raggedright
\textbf{Holm p}
\end{minipage} & \begin{minipage}[b]{\linewidth}\raggedright
\textbf{DiD {[}95\% CI{]}}
\end{minipage} & \begin{minipage}[b]{\linewidth}\raggedright
\textbf{p}
\end{minipage} & \begin{minipage}[b]{\linewidth}\raggedright
\textbf{Holm p}
\end{minipage} \\
\end{longtable}

\emph{Note.} Values are percentages or percentage points. Rows are
ordered by DiD. Confidence intervals are unadjusted. Every Black-reply
game began with 1.e4.

\textbf{Table S7.} White first moves pooled across strength settings,
with bootstrap 95\% rank intervals at each setting.

\begin{longtable}[]{@{}
  >{\raggedright\arraybackslash}p{(\columnwidth - 12\tabcolsep) * \real{0.2400}}
  >{\raggedright\arraybackslash}p{(\columnwidth - 12\tabcolsep) * \real{0.0900}}
  >{\raggedright\arraybackslash}p{(\columnwidth - 12\tabcolsep) * \real{0.1799}}
  >{\raggedright\arraybackslash}p{(\columnwidth - 12\tabcolsep) * \real{0.1299}}
  >{\raggedright\arraybackslash}p{(\columnwidth - 12\tabcolsep) * \real{0.1200}}
  >{\raggedright\arraybackslash}p{(\columnwidth - 12\tabcolsep) * \real{0.1200}}
  >{\raggedright\arraybackslash}p{(\columnwidth - 12\tabcolsep) * \real{0.1203}}@{}}
\toprule\noalign{}
\begin{minipage}[b]{\linewidth}\raggedright
\textbf{Move}
\end{minipage} & \begin{minipage}[b]{\linewidth}\raggedright
\textbf{Wins / 870}
\end{minipage} & \begin{minipage}[b]{\linewidth}\raggedright
\textbf{Win \% {[}95\% CI{]}}
\end{minipage} & \begin{minipage}[b]{\linewidth}\raggedright
\textbf{Deviation from setting mean (pp)}
\end{minipage} & \begin{minipage}[b]{\linewidth}\raggedright
\textbf{Rank 1320}
\end{minipage} & \begin{minipage}[b]{\linewidth}\raggedright
\textbf{Rank 2000}
\end{minipage} & \begin{minipage}[b]{\linewidth}\raggedright
\textbf{Rank 2500}
\end{minipage} \\
\begin{minipage}[b]{\linewidth}\raggedright
1.d4 (Queen\textquotesingle s Pawn)
\end{minipage} & \begin{minipage}[b]{\linewidth}\raggedright
425
\end{minipage} & \begin{minipage}[b]{\linewidth}\raggedright
48.9 {[}45.5, 52.2{]}
\end{minipage} & \begin{minipage}[b]{\linewidth}\raggedright
+2.6
\end{minipage} & \begin{minipage}[b]{\linewidth}\raggedright
1--9
\end{minipage} & \begin{minipage}[b]{\linewidth}\raggedright
1--4
\end{minipage} & \begin{minipage}[b]{\linewidth}\raggedright
2--10
\end{minipage} \\
\end{longtable}

\emph{Note.} Rank intervals come from 20,000 parametric bootstrap
replicates centered on the observed proportions. They therefore give an
optimistic picture of rank certainty (Section 3.3).

\textbf{Table S8.} Black first replies pooled across strength settings,
with bootstrap 95\% rank intervals at each setting.

\begin{longtable}[]{@{}
  >{\raggedright\arraybackslash}p{(\columnwidth - 12\tabcolsep) * \real{0.2400}}
  >{\raggedright\arraybackslash}p{(\columnwidth - 12\tabcolsep) * \real{0.0900}}
  >{\raggedright\arraybackslash}p{(\columnwidth - 12\tabcolsep) * \real{0.1799}}
  >{\raggedright\arraybackslash}p{(\columnwidth - 12\tabcolsep) * \real{0.1299}}
  >{\raggedright\arraybackslash}p{(\columnwidth - 12\tabcolsep) * \real{0.1200}}
  >{\raggedright\arraybackslash}p{(\columnwidth - 12\tabcolsep) * \real{0.1200}}
  >{\raggedright\arraybackslash}p{(\columnwidth - 12\tabcolsep) * \real{0.1203}}@{}}
\toprule\noalign{}
\begin{minipage}[b]{\linewidth}\raggedright
\textbf{Reply}
\end{minipage} & \begin{minipage}[b]{\linewidth}\raggedright
\textbf{Wins / 870}
\end{minipage} & \begin{minipage}[b]{\linewidth}\raggedright
\textbf{Win \% {[}95\% CI{]}}
\end{minipage} & \begin{minipage}[b]{\linewidth}\raggedright
\textbf{Deviation from setting mean (pp)}
\end{minipage} & \begin{minipage}[b]{\linewidth}\raggedright
\textbf{Rank 1320}
\end{minipage} & \begin{minipage}[b]{\linewidth}\raggedright
\textbf{Rank 2000}
\end{minipage} & \begin{minipage}[b]{\linewidth}\raggedright
\textbf{Rank 2500}
\end{minipage} \\
\begin{minipage}[b]{\linewidth}\raggedright
1...c5 (Sicilian)
\end{minipage} & \begin{minipage}[b]{\linewidth}\raggedright
383
\end{minipage} & \begin{minipage}[b]{\linewidth}\raggedright
44.0 {[}40.8, 47.3{]}
\end{minipage} & \begin{minipage}[b]{\linewidth}\raggedright
+3.2
\end{minipage} & \begin{minipage}[b]{\linewidth}\raggedright
1--10
\end{minipage} & \begin{minipage}[b]{\linewidth}\raggedright
1--7
\end{minipage} & \begin{minipage}[b]{\linewidth}\raggedright
1--5
\end{minipage} \\
\end{longtable}

\emph{Note.} Rank intervals come from 20,000 parametric bootstrap
replicates centered on the observed proportions. They therefore give an
optimistic picture of rank certainty (Section 3.3). Every Black-reply
game began with 1.e4.

\textbf{Table S9.} White\textquotesingle s first-move advantage (FMA) in
Study 1, with assumption-free bounds and expected-score bounds by
setting.

%

\emph{Note.} Point estimates, run counts, t tests, and implied expected
scores assume equal draw rates in the two sets. The bounds, derived in
Appendix A, do not. At 2000, the draw-dependent quantities use the 28
runs with a recorded draw rate.

\textbf{Table S10.} Dispersion of run-level rates relative to binomial
expectation.

%

\emph{Note.} Expected variance is p(1 − p)/n, with n = 100 for win rates
and n = 200 for draw rates. Ratios above one indicate more run-to-run
variation than independent sampling would predict.

\textbf{Table S11.} Setting-level outcomes in Study 1 and in the Study 2
replication.

%

\emph{Note.} Study 2 contributes 300 games per set and setting (600 for
draws). Intervals use a normal approximation.

\textbf{Table S12.} Study 2 termination conditions, draw rates by set,
and mean game length.

%

\emph{Note.} Percentages are based on the 600 games at each nominal
setting. Games reaching the 300-ply cap were scored as draws.

\textbf{Table S13.} Study 3 matches between nominal strength settings,
with colors alternated.

%

\emph{Note.} Estimated gaps use the logistic Elo formula. A perfect
score gives an unbounded estimate. Nominal gaps use the 1350 floor for
the lowest setting.

\textbf{Table S14.} Study 4 outcomes for 1.d3 compared with the other
nine White first moves, in games played after the hypothesis was fixed.

%

\emph{Note.} Each White move has 90 games per setting (270 in total for
1.d3 and 2,430 for the other moves). Expected score counts a draw as
half a win. Prespecified one-sided Mantel--Haenszel test: common odds
ratio 0.90, 95\% CI {[}0.69, 1.16{]}, p = .208. Test of equal odds
ratios across settings: p = .650.

\textbf{Table S15.} Study 5 Black replies selected by the opening-survey
rule.

%

\emph{Note.} Black played at nominal 2000 throughout. The scores are
shrinkage-adjusted training estimates used for selection, not estimates
from the independent evaluation. All six selected cells had 290
completed outcomes.

\textbf{Table S16.} Study 5 differences in Black expected score in the
independent evaluation.

%

\emph{Note.} Negative estimates favor letting Stockfish choose freely.
Intervals and p-values use block-level normal approximations. Holm
adjustment applies only to the two prespecified learned-policy
contrasts.

\textbf{Table S17.} Study 5 difference between the learned policy and
freely choosing Stockfish, by White first move.

%

\emph{Note.} Game-level normal-approximation intervals are unadjusted.
Censored games are excluded rather than scored as draws.

\textbf{Table S18.} Study 6 Black expected score by forced variation and
White setting.

\begin{longtable}[]{@{}
  >{\raggedright\arraybackslash}p{(\columnwidth - 6\tabcolsep) * \real{0.2200}}
  >{\raggedright\arraybackslash}p{(\columnwidth - 6\tabcolsep) * \real{0.2200}}
  >{\raggedright\arraybackslash}p{(\columnwidth - 6\tabcolsep) * \real{0.2799}}
  >{\raggedright\arraybackslash}p{(\columnwidth - 6\tabcolsep) * \real{0.2801}}@{}}
\toprule\noalign{}
\begin{minipage}[b]{\linewidth}\raggedright
\textbf{Family}
\end{minipage} & \begin{minipage}[b]{\linewidth}\raggedright
\textbf{Variation}
\end{minipage} & \begin{minipage}[b]{\linewidth}\raggedright
\textbf{Against White 1800 (\%) {[}95\% CI{]}}
\end{minipage} & \begin{minipage}[b]{\linewidth}\raggedright
\textbf{Against White 2000 (\%) {[}95\% CI{]}}
\end{minipage} \\
\begin{minipage}[b]{\linewidth}\raggedright
Sicilian
\end{minipage} & \begin{minipage}[b]{\linewidth}\raggedright
Najdorf
\end{minipage} & \begin{minipage}[b]{\linewidth}\raggedright
66.6 {[}61.7, 71.6{]}
\end{minipage} & \begin{minipage}[b]{\linewidth}\raggedright
52.8 {[}47.5, 58.0{]}
\end{minipage} \\
\begin{minipage}[b]{\linewidth}\raggedright
Sicilian
\end{minipage} & \begin{minipage}[b]{\linewidth}\raggedright
Dragon
\end{minipage} & \begin{minipage}[b]{\linewidth}\raggedright
65.5 {[}60.7, 70.4{]}
\end{minipage} & \begin{minipage}[b]{\linewidth}\raggedright
46.0 {[}40.9, 51.2{]}
\end{minipage} \\
\begin{minipage}[b]{\linewidth}\raggedright
Sicilian
\end{minipage} & \begin{minipage}[b]{\linewidth}\raggedright
Classical
\end{minipage} & \begin{minipage}[b]{\linewidth}\raggedright
61.4 {[}56.3, 66.6{]}
\end{minipage} & \begin{minipage}[b]{\linewidth}\raggedright
45.3 {[}40.3, 50.4{]}
\end{minipage} \\
\begin{minipage}[b]{\linewidth}\raggedright
Sicilian
\end{minipage} & \begin{minipage}[b]{\linewidth}\raggedright
Scheveningen
\end{minipage} & \begin{minipage}[b]{\linewidth}\raggedright
60.7 {[}55.5, 65.9{]}
\end{minipage} & \begin{minipage}[b]{\linewidth}\raggedright
39.1 {[}33.9, 44.3{]}
\end{minipage} \\
\end{longtable}

\emph{Note.} Black played at nominal 2000. Each cell contains 290
scheduled games, and six cells have 289 completed outcomes because one
game was censored. Intervals are descriptive and not simultaneous.

\textbf{Table S19.} Study 6 within-family contrasts that survived Holm
adjustment across 42 comparisons.

\begin{longtable}[]{@{}
  >{\raggedright\arraybackslash}p{(\columnwidth - 8\tabcolsep) * \real{0.1400}}
  >{\raggedright\arraybackslash}p{(\columnwidth - 8\tabcolsep) * \real{0.4000}}
  >{\raggedright\arraybackslash}p{(\columnwidth - 8\tabcolsep) * \real{0.1599}}
  >{\raggedright\arraybackslash}p{(\columnwidth - 8\tabcolsep) * \real{0.1500}}
  >{\raggedright\arraybackslash}p{(\columnwidth - 8\tabcolsep) * \real{0.1501}}@{}}
\toprule\noalign{}
\begin{minipage}[b]{\linewidth}\raggedright
\textbf{White setting}
\end{minipage} & \begin{minipage}[b]{\linewidth}\raggedright
\textbf{Contrast}
\end{minipage} & \begin{minipage}[b]{\linewidth}\raggedright
\textbf{Difference (pp)}
\end{minipage} & \begin{minipage}[b]{\linewidth}\raggedright
\textbf{Holm-adjusted p}
\end{minipage} & \begin{minipage}[b]{\linewidth}\raggedright
\textbf{99.88\% CI}
\end{minipage} \\
\begin{minipage}[b]{\linewidth}\raggedright
2000
\end{minipage} & \begin{minipage}[b]{\linewidth}\raggedright
Sicilian Najdorf minus Sicilian Scheveningen
\end{minipage} & \begin{minipage}[b]{\linewidth}\raggedright
+13.68
\end{minipage} & \begin{minipage}[b]{\linewidth}\raggedright
.0013
\end{minipage} & \begin{minipage}[b]{\linewidth}\raggedright
{[}+3.73, +23.62{]}
\end{minipage} \\
\begin{minipage}[b]{\linewidth}\raggedright
1800
\end{minipage} & \begin{minipage}[b]{\linewidth}\raggedright
Queen\textquotesingle s Gambit Declined minus Slav
\end{minipage} & \begin{minipage}[b]{\linewidth}\raggedright
+13.28
\end{minipage} & \begin{minipage}[b]{\linewidth}\raggedright
.0053
\end{minipage} & \begin{minipage}[b]{\linewidth}\raggedright
{[}+2.49, +24.07{]}
\end{minipage} \\
\midrule\noalign{}
\endhead
\bottomrule\noalign{}
\endlastfoot
\end{longtable}

\emph{Note.} The paired tests use 29 block differences with 28 degrees
of freedom. The intervals use a Bonferroni level of .05/42. Both
contrasts are exploratory.

\textbf{Table S20.} Connect Four positive control: first-player win
percentage by forced opening column (29 runs, 290 games per column,
weakest bot on both sides).

%

\emph{Note.} Center-column contrast: 22.5 points (95\% CI {[}16.5,
28.5{]}). Homogeneity across columns: χ²(6) = 64.47, p \textless{} .001.
Ranking reliability 0.91. The center column had the highest score,
counting ties, in 20 of 29 runs.

\subsection{Supplementary Figures}\label{supplementary-figures}

\includegraphics[width=6in,height=2.52083in]{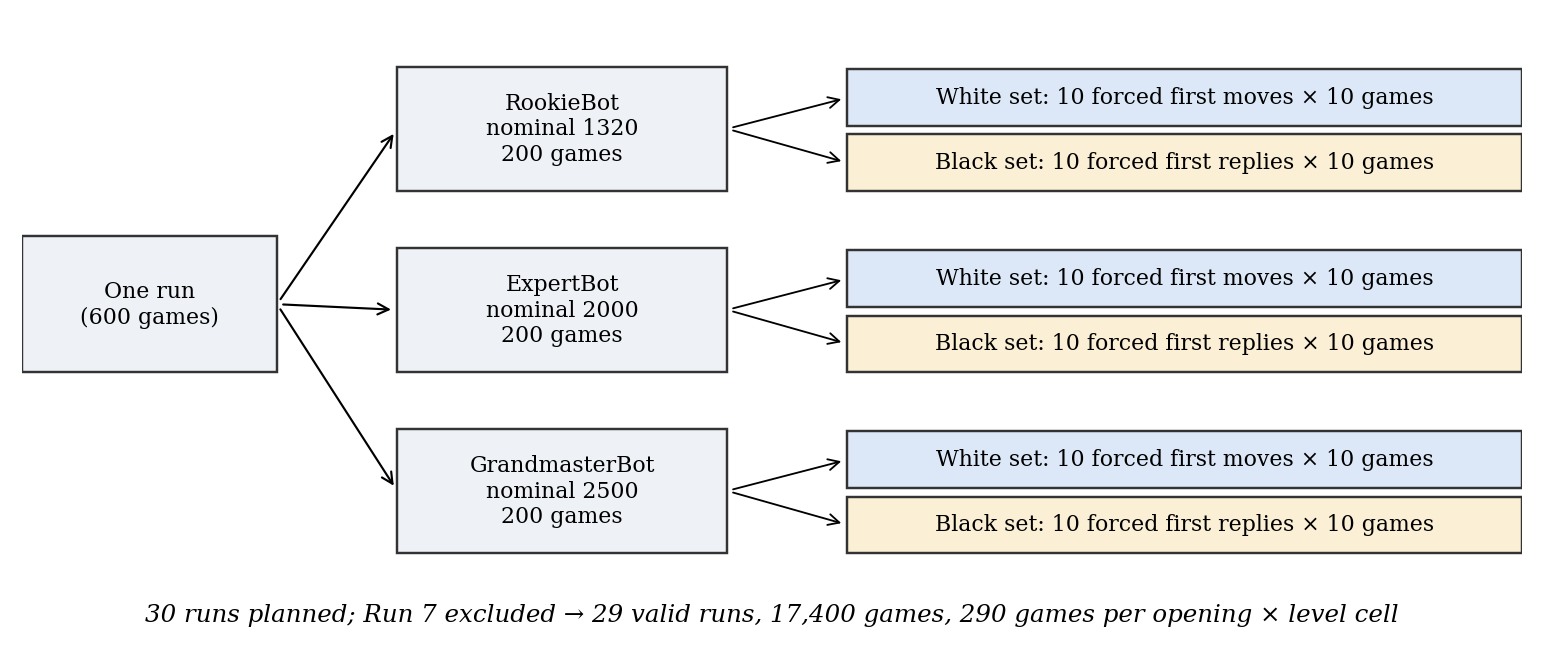}

\textbf{Figure S1.} Structure of one Study 1 simulation run. Each of the
three strength settings played 100 games with a forced White first move
and 100 games with a forced Black first reply, ten games per forced
move. Both sides of each game used the same setting.

\includegraphics[width=6in,height=3.84375in]{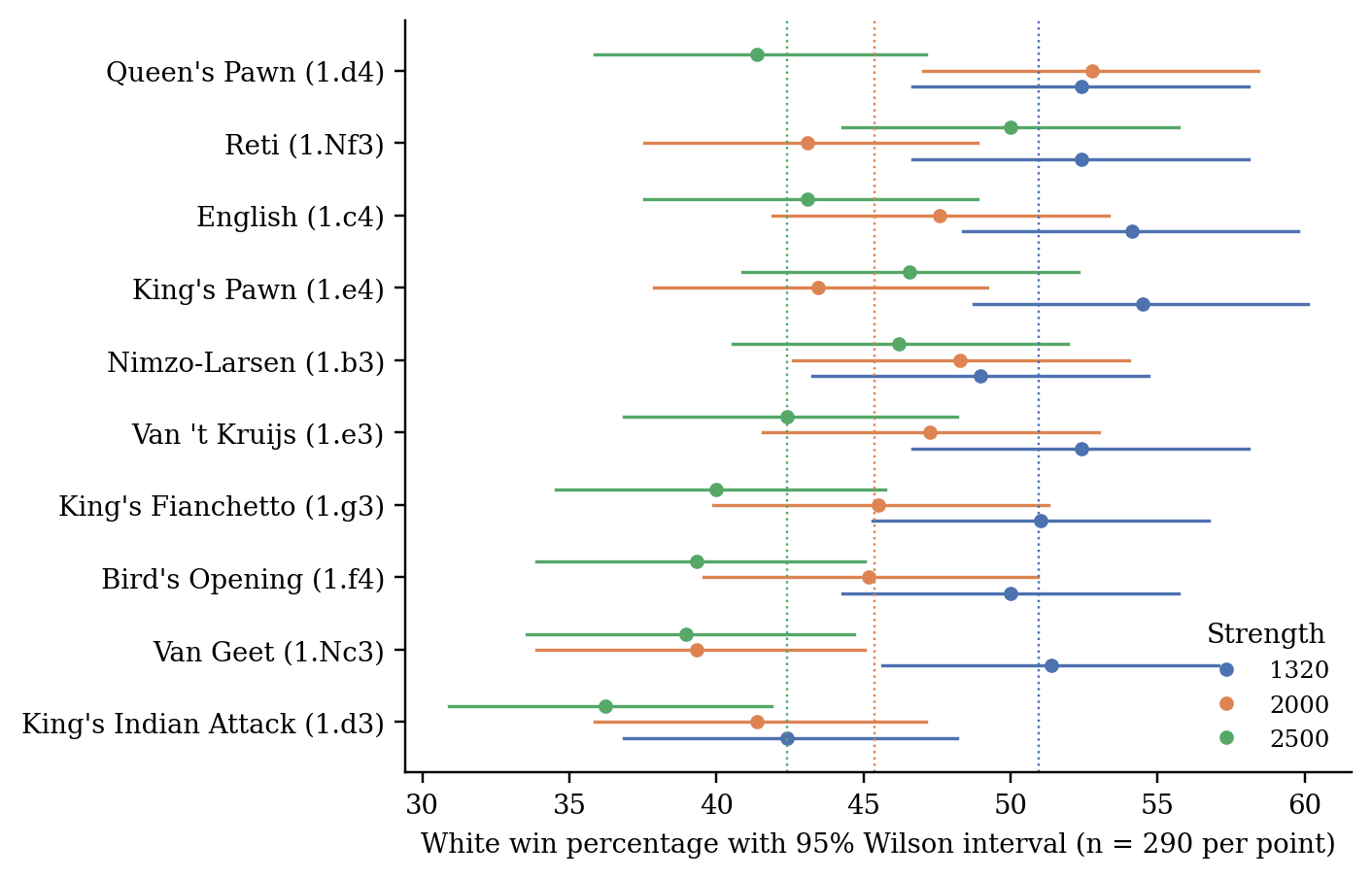}

\textbf{Figure S2.} White win percentage by forced first move and
strength setting, with 95\% Wilson intervals (n = 290 games per point).
Dotted vertical lines mark the setting means. Moves are ordered by win
rate pooled across settings.

\includegraphics[width=6in,height=3.83333in]{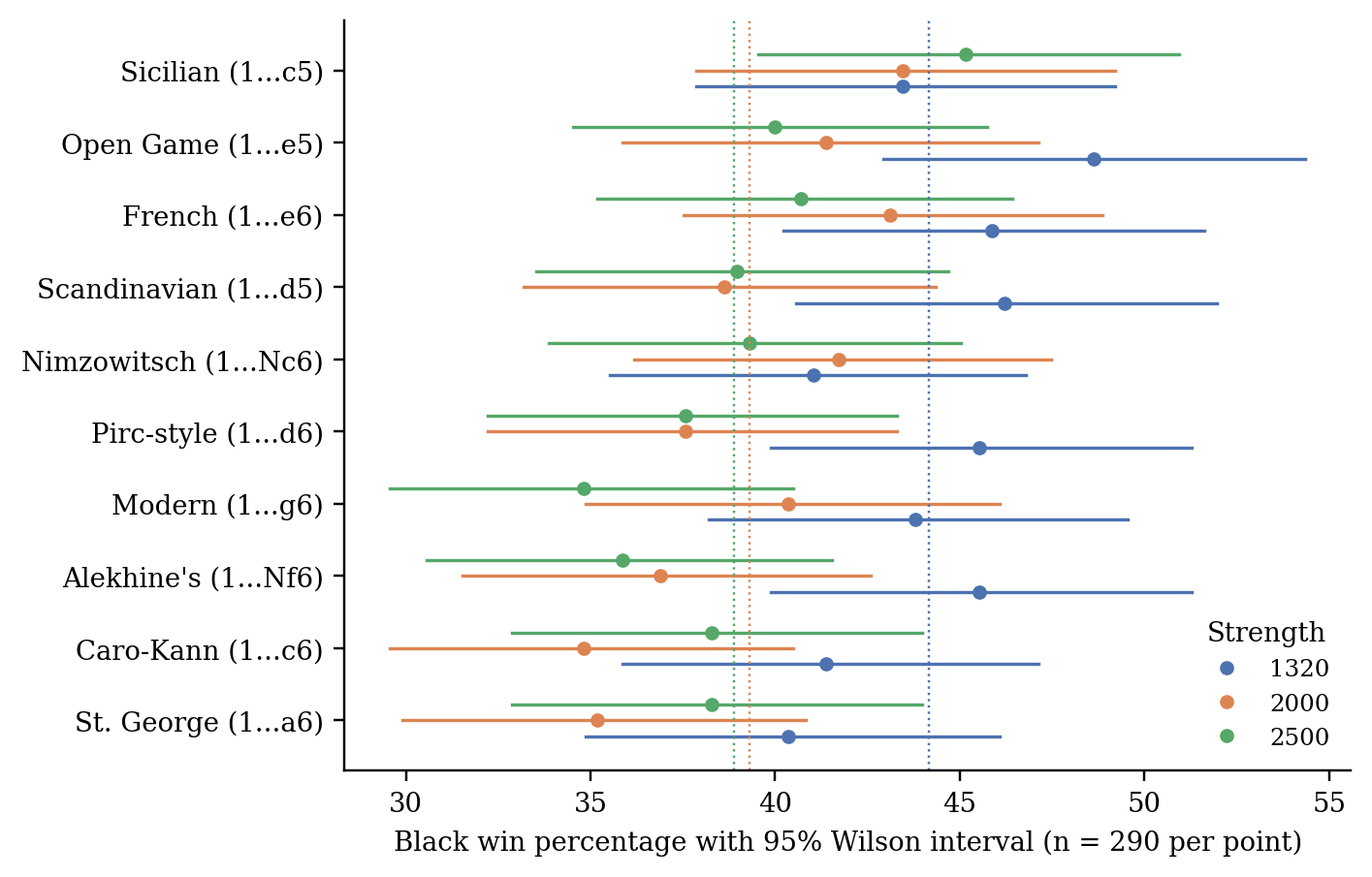}

\textbf{Figure S3.} Black win percentage by forced first reply and
strength setting, with 95\% Wilson intervals (n = 290 games per point).
Dotted vertical lines mark the setting means.

\includegraphics[width=6in,height=3.38542in]{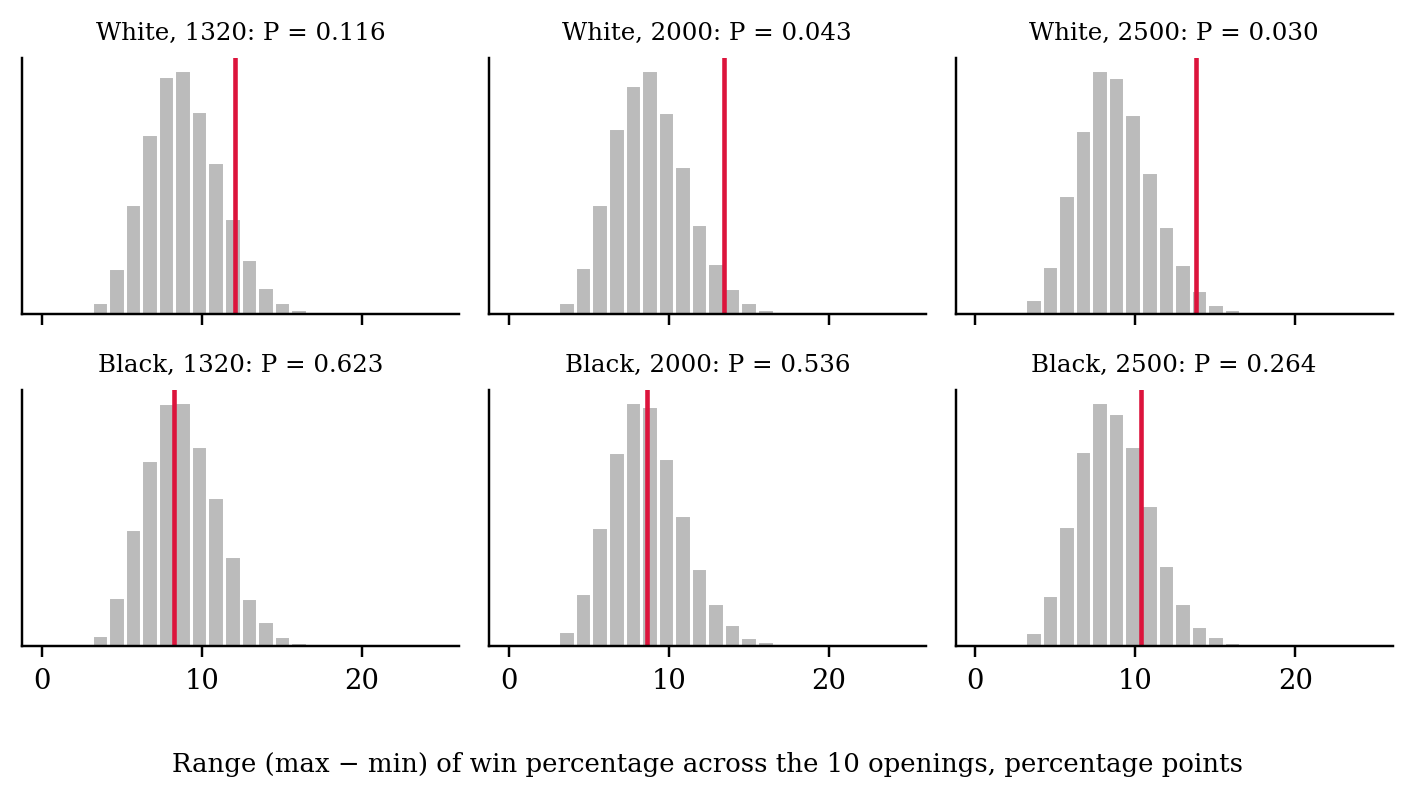}

\textbf{Figure S4.} Observed range of win percentages across the ten
openings (red line), compared with the simulated distribution under
equal win probabilities (100,000 simulations). Each
panel\textquotesingle s p-value is the proportion of simulated ranges at
least as large as the observed range.

\includegraphics[width=6in,height=3.85417in]{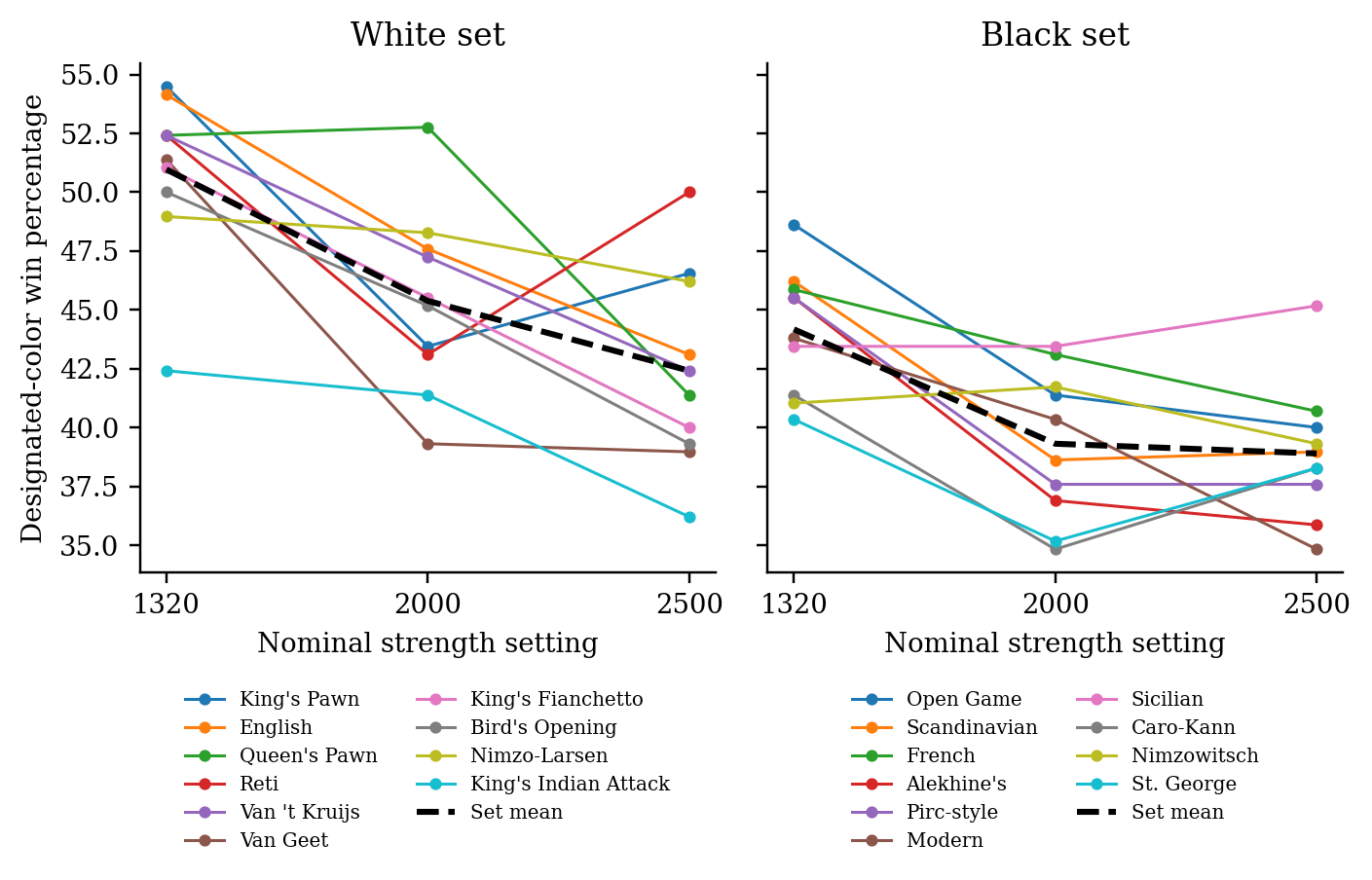}

\textbf{Figure S5.} Designated-color win percentage of each opening
across strength settings. Dashed black lines show the mean of the ten
openings for that color. With no interaction, the lines would be
parallel on the logit scale apart from sampling error.

\includegraphics[width=6in,height=3.10417in]{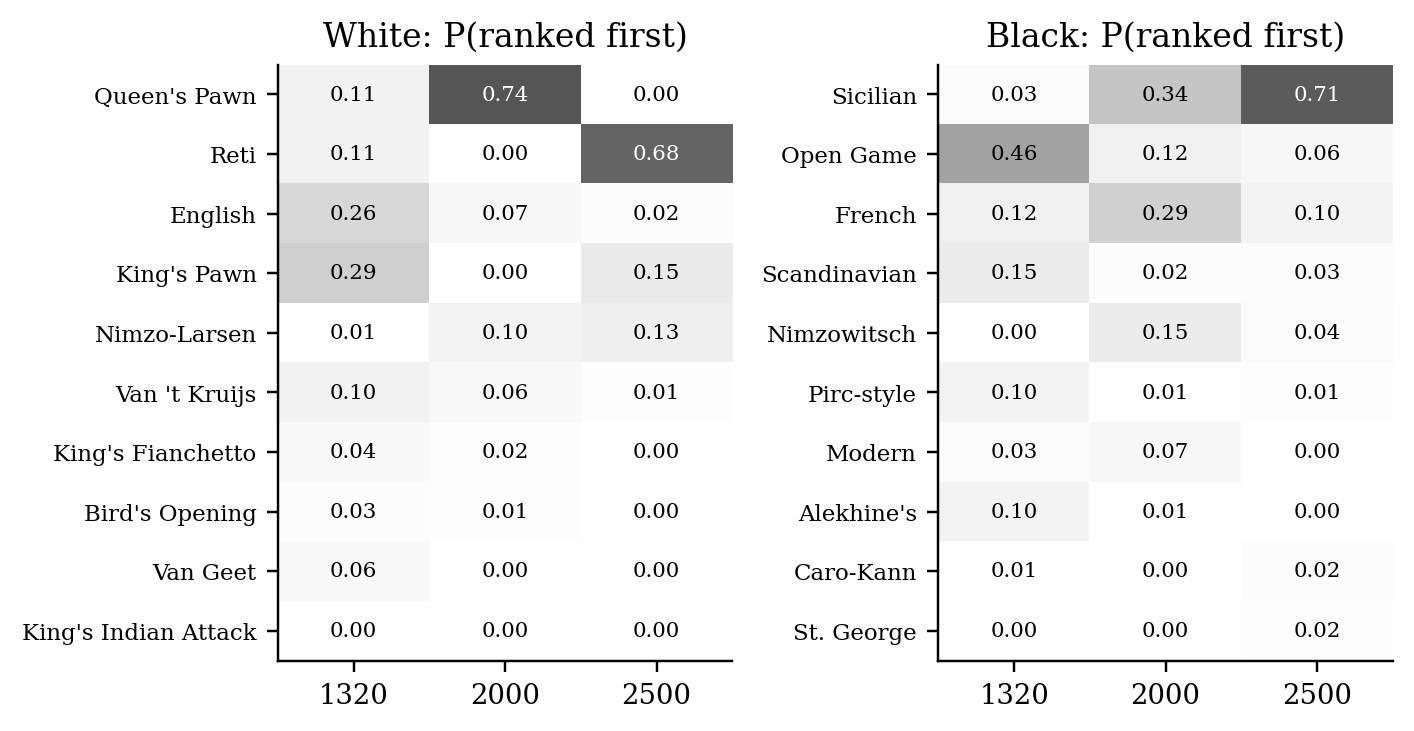}

\textbf{Figure S6.} Bootstrap probability that each opening ranks first
at each strength setting (20,000 replicates centered on observed
proportions). Because the resampling is centered on observed values,
these probabilities overstate certainty in the observed leaders.

\includegraphics[width=6in,height=2.91667in]{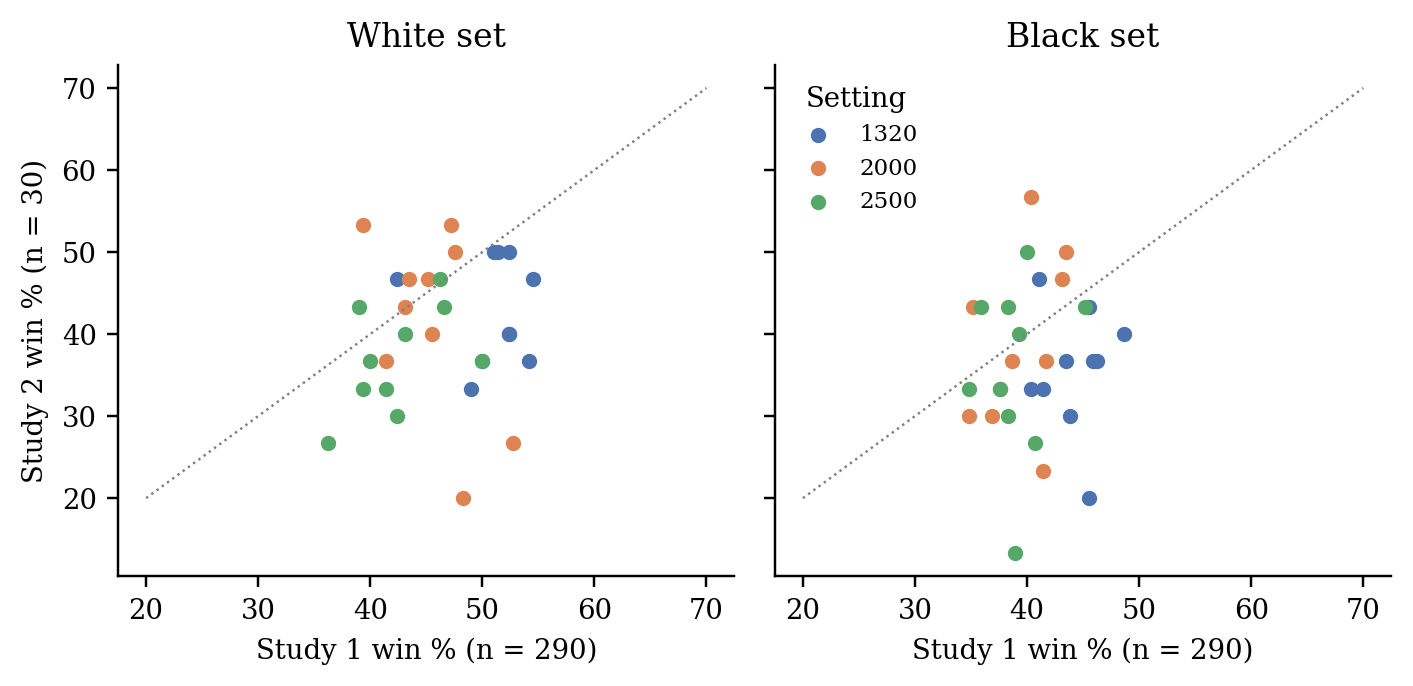}

\textbf{Figure S7.} Designated-color win percentage in each of the 60
opening-by-setting cells in Study 1 (290 games per cell) and Study 2 (30
games per cell). The diagonal marks equality. Scatter around it is
expected given the small Study 2 cells.

\end{document}